\documentclass[aps]{revtex4}
\usepackage{graphicx}% Include figure files
\usepackage{dcolumn}% Align table columns on decimal point
\usepackage{bm}% bold math
\usepackage{amssymb}
\usepackage{rotate}
\usepackage{epsfig}
\newcommand \be  {\begin{equation}}
\newcommand \bea {\begin{eqnarray} \nonumber }
\newcommand \ee  {\end{equation}}
\newcommand \eea {\end{eqnarray}}
\begin{document}
\title{Crises and collective socio-economic phenomena: simple models and challenges}

\author{Jean-Philippe Bouchaud}

\affiliation{
Capital Fund Management, 23 rue de l'Universit\'e,
75007 Paris, France\\
}
\affiliation{Ecole Polytechnique, Route de Saclay,  91120 Palaiseau, France.}
\date{\today}

\begin{abstract}
Financial and economic history is strewn with bubbles and crashes, booms and busts, crises and upheavals of all sorts. Understanding the origin of these events is arguably one of the most important 
problems in economic theory. In this paper, we review recent efforts to include {\it heterogeneities and interactions} in models of decision. We argue that the so-called Random Field Ising model ({\sc rfim})  provides a unifying framework to account for many collective socio-economic phenomena that lead to sudden ruptures and crises. We discuss different models that can capture potentially destabilising 
self-referential feedback loops, induced either by {\it herding}, i.e.  reference to peers,  or {\it trending}, i.e. reference to the past, and that account for some of the phenomenology missing in the standard models. We discuss some empirically testable predictions of these models, for example robust signatures of {\sc rfim}-like herding effects, or the logarithmic decay of spatial correlations of voting patterns. One of the most striking result, inspired by statistical physics methods, is that Adam Smith's invisible hand can fail badly at solving simple coordination problems. We also insist on the issue of time-scales, that can be extremely long in some cases, and prevent socially optimal equilibria from being reached. As a theoretical challenge, the study of so-called ``detailed-balance'' violating decision rules is needed to decide whether conclusions based on current models (that all assume detailed-balance) are indeed robust and generic. 
\end{abstract}

\pacs{Valid PACS appear here}% PACS, the Physics and Astronomy

\maketitle
\tableofcontents

\section{Introduction}

Traditional economics treats the aggregate behaviour of a whole
population through a ``representative agent'' approach, where the 
heterogeneous preferences of individual agents are replaced by average 
preferences curve. Agents are supposed to rationaly determine their action in isolation,
with no reference whatsoever to the decision of others; in other words, {\it interactions}
between agents are usually neglected. The need to account for interactions seems however quite compelling: 
imitation and social pressure effects must be responsible for the appearance of trends, 
fads and fashion or mass panic, that would be very difficult to understand if agents were really insensitive to the 
behaviour of others. \footnote{As a recent anecdotal, but pittoresque piece of evidence: the explosion of ``love-locks'' on 
the {\it Pont des Arts} in Paris since 2008.} 
Financial markets -- and for that matter whole economies -- are prone to crises 
and exhibit sudden discontinuities and crashes, even in the absence of any violent exogenous event. The 
idea that some crises are endogenously generated dates back to MacKay \cite{MacKay}, Keynes \cite{Keynes} and Minsky \cite{Minsky}, 
with many more recent apostles -- see e.g. 
\cite{Soros,Akerlof,Kirman_Book,Sornette1,Sornette2,Risk}. Collective effects can be the key to the success of a brand, a book or a movie \cite{Rodgers}, 
or else of a new technology or a vaccination campaign. But they can also be detrimental and do lead to major catastrophes when imitation cascades 
are based on unreliable information or dangerous ideas, and when social pressure supersedes rational thinking.
In any case, social imitation often leads to distortion and exaggeration, i.e. a decoupling
between the cause and the effect, which in turn generates inequalities and concentration (or ``condensation'') phenomena, see sections \ref{sect4D}, \ref{sect6A}.
These effects may grow ever stronger in our Internet and social network dominated societies. 

From a theoretical point of view, interactions may indeed lead to an aggregate behaviour that is completely
different from that implied by standard representative agent approaches \cite{Kirman_RA}. Catastrophic events and discontinuities 
can occur at the macro level, when the behaviour of independent agents would be
perfectly smooth and featureless. \footnote{The converse is in fact also true: discontinuous behaviour at the agent level may end up being smoothed out at the macro level!}
Models of collective behaviour that illustrate those ``surprises'' on the 
way from micro to macro are of course familiar to statistical physicists -- with Van der Waals 
and Weiss as early heroes \cite{Sethna_Book}. Similar ideas started to develop in quantitative sociology in 
the seventies, with models introduced by Schelling (1971, 1973) \cite{Schelling} and Granovetter (1978) \cite{Granovetter} that are 
formally very close to spin models of magnets (see below, section \ref{sect3A}). In fact, the analogy between magnetic polarization 
and opinion polarization was noted and formalized slightly earlier by a physicist (Weidlich) in 1971 \cite{Weidlich_1} who
coined the word ``Sociodynamics'', and slightly later by Galam, Gefen \& Shapir (1982) \cite{Galam_0}, in a manifesto for
``Sociophysics''.  

As it is well known, interactions can lead to multiple equilibria and ergodicity breaking, which in turn lead to
very interesting effects like ``punctuated equilibrium'', hysteresis (or path dependence) and aging. These theoretical scenarios offer enticing 
metaphors for understanding economic systems as well, in particular the emergence of trust, norms and conventions. After all, a prosperous economy 
can only be achieved through efficient coordination and collaboration. Although anticipated by Keynes and pioneered by F\"ollmer in 1974 (again in the context of spin models! \cite{Follmer}), 
these ideas only really permeated into economics in the early nineties, partly nurtured by the Santa Fe meetings ``{\it The Economy As an Evolving Complex System}'' \cite{SantaFe1,SantaFe2}. 
As Kirman (1992-93) was forcefully arguing against the representative agent framework \cite{Kirman_RA} and proposing his  `ant' recruitment model \cite{Kirman_ant}, 
Brock and Durlauf wrote a series of remarkable papers \cite{Durlauf,Brock_Hommes,Brock} using statistical mechanics methods to elicit the importance
of social interactions in economics, emphasizing that {\it the utility or payoff an individual receives depends directly on the choices of others.} \footnote{See also 
the earlier insightful paper by Becker \cite{Becker_restaurant}.}
Although not yet fully mainstream, the field of ``social dynamics'' is now well appreciated 
 \footnote{The paper {\it Discrete choice with social interactions} by Brock and Durlauf \cite{Brock} has 976 Google scholar citations at the time of writing.} 
(see the books \cite{Becker_Book, Durlauf_Book} for enlightening discussions,  \cite{BHW,Chamley} and references therein for another strand of literature 
on ``herding'' effects; see also \cite{Orlean}). 
In parallel, physicists have actively contributed to the field, with a variety of models for coordination, 
contagion, opinion formation, segregation, herding, etc. -- see e.g. \cite{MG,Fortunato,Buchanan,Ball} for recent overviews of these endeavours.

The present paper is intended to be a personal, somewhat informal review, centred around the use of Random Field Ising models ({\sc rfim}) as a generic model for 
crashes, opinion swings and discontinuities in a socio-economic context. The {\sc rfim}  provides 
a unifying framework and an enticing ``story'' for many disparate collective phenomena, including overnight trust evaporation 
such as that which occured after Lehman's bankruptcy in 2008, or the persistence of cultural traits or habits, even when damaging for the community. \footnote{Think 
for example of the tax-evasion culture in some countries, that has recently become an acute issue.}
Interestingly, some of the predictions are robust and can be quantitatively tested against data. We then discuss some theoretical 
limitations of the simplest (mean-field) version of the model. Three issues are, to our way of thinking, particularly relevant. One is the
importance of history and memory in the preference of individual agents, in the absence of any interaction effects (this is called
habit formation or ``stickiness'') -- see section \ref{sect6A}. Another is the use of decision making rules (or selection probabilities) that have become standard 
in the above cited literature, and that amount to assuming what is called {\it detailed-balance} in the physics jargon -- see section \ref{sect5B} for a precise 
definition. How much {\it non} detailed-balance rules
would affect the qualitative features is very much an open problem (section \ref{sect5B}). The third one concerns spatial effects, that should 
play a major role in some cases, as illustrated by election turnout statistics (section \ref{sect4C}). Finally, we cursorily discuss alternative (or rather,
complementary) models of crashes and regime shifts based on feedback loops between past events and current decisions (sections \ref{sect6B}, \ref{sect6C}).  

Many ideas in this paper have been ``long and widely known''. However, they are only well known to those who know them well. Our experience is 
that even simple ideas may have a hard time permeating into other communities (the adoption models presented below could even be used to explain why!
\cite{Helbing_boost}). Sometimes, a detailed translation helps. We hope that the present paper can serve this purpose and be a helpful exposition, in particular for
students in physics and in economics interested by the issue of crises and collective phenomena.

\section{A generic binary choice model: the Random Field Ising Model}

\subsection{Set up of the model}
\label{sect1A}

The so-called `Random Field Ising Model' ({\sc rfim}) was proposed to model hysteresis loops and other anomalous properties 
of disordered magnets \cite{Sethna1,Sethna2}. The hysteresis loop problem is an example of a collective dynamics of flips (the individual magnetic spins) under the
influence of a slowly evolving external solicitation. Depending on the parameters, the flips may take place smoothly, 
or be organized in intermittent bursts, or `avalanches', leading to a specific acoustic pattern called Barkhausen noise.
But because the model is so generic, it can be used as a simplified model for a host of other physical situations as
well, such as earthquakes, fracture in disordered materials, failures in power grids, etc. These systems respond to the
driving in a series of avalanches spanning a broad range of scales, which Sethna, Dahmen and Myers (2001) 
proposed to call ``crackling noise'' \cite{Sethna_nature}.

As was originally proposed in \cite{Galam_0,Galam_1}, the model can also be transposed to represent binary decision situations under social pressure, 
and influenced by some global information (such as the price of a product or the quality of a technology) 
or by {\it zeitgeist}. This transposition was recently discussed in several socio-economical contexts in, e.g. 
\cite{QF,Nadal1,Nadal2,Nadal3,Michard,Harras} (see \cite{Vives,Marsili_Kirman,Lorenz,Holyst} for variations on the theme).  The model has a rich
phenomenology. In particular, discontinuities appear in aggregate quantities when imitation effects exceed a certain threshold, even 
if the external solicitation varies smoothly with time. In this situation, the aggregate demand is a multi-valued function of the price \cite{Nadal1}. 
Below the threshold, on the other hand, the behaviour of ``demand'', or of the average opinion, is smooth, 
but the natural trend can be substantially amplified and accelerated by peer pressure, possibly leading to sudden crises. 

We will assume that each agent $i$ is confronted to a binary choice, the outcome of which is denoted by $S_i = -1, 1$.
This binary choice can be to vote or not to vote, or to vote yes or no in a referendum; it can be to
buy or not to buy a certain good \cite{Nadal1,Nadal2,Nadal3}, to evade tax or not, to clap or to stop clapping at the end
of concerts, etc. \cite{Michard}. Textbook examples in sociology are to attend or not to attend a seminar \cite{Schelling}, 
to join or not to join a riot \cite{Granovetter} or a strike \cite{Galam_0}, to go or not to go to the crowded El Farol bar \cite{Arthur,MG}. 
Of course, examples can be multiplied {\it ad libitum}. We wish to 
add two more in view of their importance: one is to trust or not to trust. This encompasses a large number of situations that recently made the 
news: credit markets and credit ratings; trust in future economic prospects; trust in the validity of the models used to value derivative products \cite{Marsili_Kirman}; 
trust in the value of money itself, etc. The second is adopting or not an environment friendly technology (i.e. solar heating) which is individually
costly but collectively valuable. This is closely related to the famous ``tragedy of the commons'' \cite{Hardin} wherein individual short term interest
is in conflict with long term collective welfare. 

It is reasonable to assume that the decision of agent $i$ depends on three distinct factors: 
\begin{itemize}
\item (i) his personal inclination, preference or ``willingness to pay/adopt'', measured by a real variable $f_i \in
(-\infty,+\infty)$ which we take to be time independent (but see section \ref{sect6A}) and distributed according to a certain p.d.f $\rho(f)$. 
Large positive $f$'s means a strong a priori tendency to decide $S=1$, and large 
negative $f$'s a strong bias towards $S=-1$.  
\item (ii) public information, affecting all agents equally, such as
objective information on the scope of the 
vote, the price of the product agents want to buy, the balance sheet of the banks, the advance of technology, etc.
The influence of this time dependent common factor will be called the incentive
field $F(t)$, again a real variable in $(-\infty,+\infty)$.
\item (iii) social pressure or imitation effects; each agent $i$ is
influenced by the previous decision made by a certain 
number of other agents $j$ in his ``neighbourhood'', ${\cal V}_i$. The
influence of $j$ on $i$ is taken to be 
$J_{ij} S_j$, that adds to $f_i$ and $F$. If $J_{ij} > 0$, the
decision of agent $j$ to buy (say) reinforces the 
attractiveness of the product for agent $i$, who is now more likely to
buy. This reinforcing effect can obviously 
lead to an unstable feedback loop, as discussed in section \ref{sect3B}. If on the
contrary $J_{ij} < 0$, the action of agent $j$ 
deters agent $i$ from making the same choice. This ``anti-conformist" or contrarian
tendency, although rarer in human nature, can 
sometimes exist and be relevant. For example, buying can push the price up and discourage others from buying themselves. 
When both signs of $J_{ij}$ coexist, one expects many metastable states with a very rich dynamics, see e.g. \cite{MM}.
\end{itemize}

\subsection{Decision rules}
\label{sect1B}

To sum up then, the overall perceived incentive $U_i$ of agent $i$ to choose $+1$ over $-1$ is 
\be
U_i(t)=f_i + F(t) + \sum_{j \in {\cal V}_i} J_{ij} S_j(t-1).
\ee
Now one has to specify a decision rule, given a certain value of the total
incentive. The simplest rule is:
\be\label{rfim}
S_i(t) = \mbox{sign}\left[U_i(t) - U_{th}\right],
\ee
meaning that the decision to ``buy'' is reached whenever the incentive $U_i$
reaches a certain {\it threshold value} $U_{th}$, which can be chosen without 
loss of generality to be zero. \footnote{Any other $i$-dependent value could have been chosen, since
this simply amounts to shifting the value of idiosyncratic field $f_i$.} 

However, the above decision rule is perhaps too deterministic and restrictive. 
One may imagine that agent $i$ could not pay attention to or misread the public information (which is
often intrinsically ambiguous and hard to interpret) or else have a random, time dependent component in 
his preference (a sophisticated way of saying people can be fickle). \footnote{The distinction between the 
two interpretations will be discussed again in section \ref{sect5A}.} A more general rule that is a standard 
in the choice/decision making literature (see section \ref{sect5A}) is the so-called ``logit rule'' or 
``quantal response'' which makes the decision a random variable, with probability:
\be\label{logit1}
P(S_i=+1; U_i) = \frac{1}{1 + e^{-\beta U_i}}; \qquad P(S_i=-1;U_i)=1 - P(S_i=+1 ; U_i), 
\ee
where $\beta$ is a parameter that specifies the amount of noise (or ``irrationality'') in the
decision process, and is the analogue of inverse temperature in physics. When $\beta \to 0$, 
incentives play no role and there is a random $1/2,1/2$ probability of the two decisions, whereas
when $\beta \to \infty$, one recovers the deterministic rule given by Eq. (\ref{rfim}). 

It is in fact very natural to generalize slightly the above rule to keep some memory of the 
previous decision and allow for possibly time dependent incentives:
\be\label{logit2}
P(S_i(t)=+1|S_i(t-1)=-1;U_i) =  \frac{\mu}{1 + e^{-\beta U_i}}; \quad P(S_i(t)=-1|S_i(t-1)=-1;U_i)= 1 - \frac{\mu}{1 + e^{-\beta U_i}},
\ee
and 
\be\label{logit2bis}
P(S_i(t)=-1|S_i(t-1)=+1;U_i) =  \frac{\mu}{1 + e^{\beta U_i}}; \quad P(S_i(t)=+1|S_i(t-1)=+1;U_i)= 1 - \frac{\mu}{1 + e^{\beta U_i}},
\ee
The coefficient $\mu \leq 1$ can be interpreted as a probability to reconsider one's past choice between 
$t-1$ and $t$. When $\mu=1$, all memory is lost and one recovers the previous rule Eq. (\ref{logit1}). 

These decision rules might at this stage appear as somewhat arbitrary, and guided by the fact that they will allow one to use directly Boltzmann-Gibbs 
measure and use results from statistical mechanics. We will comment on those rules below (sections \ref{sect5A},\ref{sect5B}) -- although considerable effort has been devoted 
to axiomatize decision theory \cite{Anderson_Palma}, we feel that there is indeed some leeway here that might lead to interesting problems for future research. 

\subsection{The mean-field limit}
\label{sect1C}

If all $J_{ij} > 0$, the above model is known in physics as the Random
Field Ising model at temperature $T=1/\beta$, and has been intensively studied in the last decades,
in particular at $T=0$ (see e.g. \cite{Sethna1,Sethna2,Dhar}). In physics, natural networks of 
connections are $d$-dimensional regular lattices, but other topologies, more natural in a socio-economic 
context such as the fully connected case, regular trees or random graphs have been studied as well. The qualitative
phenomenology however does not depend much on the chosen topology, nor on the distribution of idiosyncratic fields $\rho(f)$, 
although quantitative details might be sensitive to these specifications. For simplicity, and in line with many previous papers \cite{Weidlich_1,Galam_0,Weidlich_review,Brock,Nadal1,Michard,Nadal2,Nadal3} 
we will thus mostly restrict attention here to the ``mean-field'' case where $J_{ij} \equiv J/N$, $\forall i, j$ pairs. This does not mean that each agent consults all the
other ones before making his mind, but rather than the average opinion $m = N^{-1} \sum_i S_i$, or total demand, becomes public
information, and influences the behaviour of each individual agent. This is in fact a very realistic assumption in many cases: for
example, the total sales of a book, or number of viewers of a movie, is certainly an important piece of information for
the consumers. It is also thought that the evolution of the public opinion can be affected by
polls or by confidence index, i.e. by a proxy of the average opinion \cite{Nadeau,Stauffer,Lux}, and, indeed, this is the reason for certain countries banning opinion polls shortly before elections. In the case of financial markets, the price change itself can
be seen, on a coarse-grained time scale, as an indicator of the aggregate demand (although the detailed relationship between the two might 
be quite subtle, see \cite{marketimpact,Toth}). While this mean-field description may be very reasonable in certain settings, one expects that network/local 
effects may be relevant in other cases, for which the neighbourhood of agents is restricted to a more specific community. This locality might actually 
change quite a bit the features of the model (see section \ref{sect3B}). 

We now turn to an analysis of the {\sc rfim}  model in different limits, and try to summarize the salient results and their interpretation in a socio-economic 
context.

\section{The case of a homogeneous population: the Ising-Weidlich model} 

\subsection{Dynamical equations and equilibrium}
\label{sect2A}

Let us start by analysing the case where $f_i \equiv f$, $\forall i$. Up to a shift in the public information field $F$, one can always set $f=0$. 
There are $N \gg 1$ agents, and at time $t$, the fraction of them who have chosen the $S=+1$ alternative is $\phi=N_+/N$. In the mean-field limit $J_{ij}
\equiv J/N$, the total incentive to choose $+1$ over $-1$ is simply:
\be
U = - C + 2J \phi,  \quad C= J - F
\ee
If we speak of adopting individual solar heating, for example, $C$ can be interpreted as the immediate cost of installing a new heating system, whereas 
the term $2J \phi$ encapsulates both the expected cost decrease and technical improvements as more people shift to the new technology, and the social pressure 
that makes it ``politically incorrect'' to stick with hydrocarbon greedy devices. 

Now, we use the dynamical updating rules given by Eqs. (\ref{logit2}) above, and set $z=e^{\beta U}$. It is easy to show that the following probabilities hold for the evolution of 
the number of `adopters', $N_+$ between $t$ and $t+1$: \footnote{It is interesting to notice that in Kirman's ant recruitment model \cite{Kirman_ant}, one rather has $P(N_+ \to N_+ + 1) = 
P(N_+ \to N_+ -1) = \mu \phi (1-\phi)$ because change of opinions are supposed to happen during two-body ``encounters'' where one of the two convinces the second one to 
change his mind. In the present setting, encounters are not necessary.}
\be\label{FP}
P(N_+ \to N_+ + 1) = \mu \frac{z}{1+z} (1-\phi); \qquad P(N_+ \to N_+ -1) =  \mu \frac{1}{1+z} \phi; \qquad P(N_+ \to N_+) = 1 - \frac{\mu}{1+z} [z + \phi(1-z)].
\ee
From the above, one gets the evolution of the average $\langle N_+ \rangle$ and the square-average  $\langle N_+^2 \rangle$:
\be
\langle N_+ \rangle_{t+1} - \langle N_+ \rangle_t = \mu [\frac{z}{1+z} -\phi]; 
\ee
and
\be
\langle N_+^2 \rangle_{t+1} - \langle N_+^2 \rangle_t = 2 \mu \langle N_+ \rangle_t [\frac{z}{1+z} -\phi] + \mu \frac{(1-\phi)z + \phi}{1+z}
\ee
These equations were obtained in the present sociological context by Weidlich \cite{Weidlich_review}, but are of course very standard in the context of the mean-field Ising model. 

Taking $\mu = {\rm d}t$ infinitesimally small for convenience, and comparing with the predictions of the following It\^o stochastic differential equation:
\be
{\rm d}N_+ = {\cal F}(\phi) {\rm d}t + \Sigma(\phi){\rm d}W
\ee
(where ${\rm d}W$ is the usual Brownian noise) one finds, by identification:
\be
{\cal F}(\phi) = \frac{z}{1+z} -\phi; \qquad \Sigma^2(\phi)= \frac{(1-\phi)z + \phi}{1+z}
\ee
In order for this equation to be well behaved in the large $N$ limit, a change of time scale must be performed as $\tilde t = t/N$ (meaning that over a finite 
change of $\tilde t$, a finite fraction of the population has been offered the possibility to switch). This finally leads to:
\be\label{Langevin}
{\rm d}\phi ={\cal F}(\phi) {{\rm d}\tilde t} + \frac{\Sigma(\phi)}{\sqrt{N}} {\rm d}\tilde W.
\ee
From now on, we will drop the tilde. Note that the scale of the noise goes down as $1/\sqrt{N}$ with the population size. 

The ``equilibrium'' state(s) of the system are such that $\langle {\rm d}\phi \rangle=0$, i.e. $\phi^*$ such that ${\cal F}(\phi^*)=0$, leading to:
\be
\phi^* = \frac{z^*}{1+z^*}, \qquad z^* = e^{\beta(-C + 2J \phi^*)}.
\ee
Setting $m = 2 \phi - 1$, one recovers (as expected) the standard Curie-Weiss mean-field equation, re-derived by Brock and Durlauf \cite{Brock} in the present framework:
\footnote{Note that there our choice of $J$ differs by a factor $2$ from the usual convention.}
\be\label{CW}
m^* \equiv 2\phi^* - 1 = \tanh\left[\frac{\beta}{2} (J m^* + F)\right].
\ee    
The solutions of that equation are well known. When $F=0$, there is a critical value $\beta_c = 2/J$ separating a high noise regime $\beta < \beta_c$ 
where agents shift randomly between the two choices, with $\phi^* = 1/2$. As first noted by Weidlich \cite{Weidlich_1}, a spontaneous ``polarization''
of the population occurs in the low noise regime $\beta > \beta_c$, i.e. $\phi^* \neq 1/2$ even in the absence of any individually preferred choice (i.e. $F=0$). 
When $F \neq 0$, one of the two equilibria is exponentially more probable than the other, and in principle the population should be locked into 
the most likely one: $\phi^* > 1/2$ whenever $F > 0$ and $\phi^* < 1/2$ whenever $F < 0$. 
In the limit $\beta \to \infty$, the above equilibrium analysis suggests that as soon as $F > 0$ 
(i.e. when the adoption cost $C$ is less than $J$), the whole population
should move towards the ``right'' equilibrium $\phi^* \approx 1$. 

Unfortunately, the equilibrium analysis is not sufficient to draw such an optimistic conclusion. A more detailed analysis of the {\it dynamics} is needed, which reveals 
that the time needed to reach equilibrium is exponentially large in the number of agents, and as noted by Keynes, {\it in the long run, we are all dead}. 
This situation is well-known to physicists, but is perhaps not so well appreciated in other circles -- for example, it is not discussed in \cite{Brock}. 
This is the main message of the present section, which would otherwise not be much more than a textbook exercise. 

\subsection{Potential barrier and metastability} 
\label{sect2B}

Define a potential function ${\cal V}(\phi)$ such that ${\cal F}  = - d{\cal V}/d\phi$. Up to an irrelevant additive constant, one finds  
\be
{\cal V}(\phi) = \frac{\phi^2}{2} - \frac{1}{2J \beta} \ln \left(1+e^{\beta (- C + 2 J \phi)}\right).
\ee
The analysis of ${\cal V}(\phi)$ is easy in the limit $\beta \gg 1$.  Since $\phi$ is $\in\, [0,1]$, one finds that as long as $|F| < J$ the potential has two minima
situated, as expected, very close to $\phi=0$ and $\phi=1$. The difference ${\cal V}(\phi=1) - {\cal V}(\phi=0)$ is equal to $-2F$ meaning, as explained above, that the 
$\phi^* \approx 1$ equilibrium is indeed favored when $F > 0$. But there also appears an energy {\it barrier} between the two minima, situated at $\phi_{\max}=(1-F/J)/2$,
which makes it hard for the system to crossover from the ``bad'' minimum to the ``good'' one. Using the above Langevin equation, Eq. (\ref{Langevin}), one finds that the 
time $\tau$ needed for the system, starting around $\phi=0$, to reach $\phi^* \approx 1$ is given by:
\be
\tau \propto  \exp \left[A N (1 - F/J)\right],
\ee
where $A$ is a numerical factor. The important point about this formula is the presence of the factor $N(1-F/J)$ in the exponential. This means that whenever $0 < F < J$, 
the system should really be in the socially good minimum $\phi^* \approx 1$, but the time to reach it is exponentially large in the population size. In other words, it has {\it no chance} 
of ever getting there on its own for large populations. Only when $F$ reaches $J$, i.e. when the adoption cost $C$ becomes {\it zero} will the population be convinced to shift to the socially optimal 
equilibrium -- simply because $\phi^* \approx 0$ is not a minimum anymore (note indeed that $\phi_{\max} \to 0$ as $F \to J$). 

As $\beta$ decreases towards $\beta_c$ (i.e. as the noise in the decision process increases), the two minima $\phi^*_{\pm}$ get closer to each other and the barrier between then gets smaller,
until hysteresis and polarization effects disappear altogether. 

Let us summarize what happens in a situation where the cost $C$ of adopting a new technology, a new idea, or a socially beneficial attitude, decreases steadily as a function of time 
in our (so far) homogeneous population. 
When individual make near-optimal choices, the population as a whole will not adopt until $C = 0$, even if it would be optimal for the community as a whole to adopt as soon as $C < J$ -- it would 
actually do so if one was prepared to wait an exponentially long time. As a function of time, one should  see an 
extremely small number of adopters until the cost drops to zero, in which case the whole population finally shifts to the new paradigm. This is very different from the standard model 
of innovation diffusion, based on the following simple differential equation proposed by Bass in 1969 (\cite{Bass}, and for a recent illuminating review \cite{Young}):
\be
\frac{d\phi}{dt} = \mu \phi (1-\phi),
\ee
which says that the speed of adoption first increases with the number of people who have adopted. This equation leads to the famous ``S-curve'' for adoption. It predicts that the 
adoption speed is maximum when $\phi=1/2$,  at variance with the above scenario where the adoption speed abruptly increases when $\phi$ is still quite small. Note however that the
discontinuity in time should be smoothed by the speed at which information about the behaviour of others is spreading; but when applied to financial markets, for example, this time can
indeed be very short.

What are the solutions to escape from the ``bad'' equilibrium as soon as it becomes collectively profitable to move to the ``good'' one? Clearly, this is a {\it coordination problem}, 
which becomes hard to solve when $N$ is large. Apart from lowering the cost $C$, the model above
suggests at least three possibilities to make the barrier easier to cross: a) increase the variance term $\Sigma$ by coordinating the moves (instead of having independent decisions):  
this can be done through advertising campaigns, word of mouth, etc.; b) decrease $\beta$, i.e. increasing the noise, for example when the information about the true costs is ambiguous; c) 
decrease $N$, i.e. the number of people in direct interaction -- people do not adopt because nobody else adopts. In physics, the existence of 
mutually inaccessible minima with different potentials is a pathology of mean-field models that disappears when the interaction is short-ranged. In this case, the transition proceeds through
``nucleation'', i.e. droplets of the good minimum appear in space and then grow by flipping spins at the boundaries. This suggests an interesting policy solution when social pressure resists the
adoption of a beneficial practice or product: subsidize the cost locally, or make the change compulsory there, so that adoption takes place in localized spots from which it will invade the whole population. The
very same social pressure that was preventing the change will make it happen as soon as it is initiated somewhere. 

Before leaving the homogeneous Ising-Weidlich model of this section, we would like to note that the solar heating example is in fact not be the best one, because the dynamics of the model assumes that
choices are to some extent reversible, i.e. each agent can in principle flip back and forth between the two options. While this is clearly the case for opinions, tax-evasion, safe driving, trust, etc., 
it is harder to imagine that agents can easily switch between solar cells and heating oil -- at least on short enough time scales to make the above description useful. 
The model in the next section is actually better suited for such irreversible choice situations.

\section{The role of heterogeneities: scaling and avalanches}

\subsection{The zero-temperature {\sc rfim}  and classical models of adoption}
\label{sect3A}

Let us now come back to the idea that agents are not clones and have different preferences or willingness to pay $f_i$, distributed according to a certain p.d.f $\rho(f)$. 
The case where social pressure is absent, i.e. $J_{ij} \equiv 0$ is very simple to analyse when $\beta \to \infty$. Call $P_<(f)=\int_{-\infty}^f {\rm d}x \rho(x)$
the cumulative distribution of $f_i$, i.e. the probability that $f_i \leq f$. The aggregate demand, or average opinion $m$ for a given 
incentive field $F(t)=F$ is easily obtained as:
\be 
m_0 = - P_<(-F) + (1 - P_<(-F)) = 1 - 2 P_<(-F).
\ee
(The subscript $0$ means that $J = 0$ here).
As $F(t)$ increases slowly from $-\infty$ to $+\infty$, the average opinion
evolves from $-1$ to $+1$ in a way that mirrors exactly the distribution of a priori opinions in the population. 
For a generic distribution of $f_i$ (for example, Gaussian), the opinion evolves {\it smoothly} as the polarization field is
increased -- see Fig. 1, blue line. If one interprets $F$ as minus the price $p$ of a product, and $f_i$ the price at which an agent is happy to buy, 
the total demand curve $d=(m+1)/2$ is clearly equal to the number of buyers, i.e.
\be
d(t) = N (1 - P_<(p(t)) \equiv N \int_{p(t)}^\infty {\rm d}x \rho(x),
\ee
which increases as the price decreases. For a unimodal distribution, $d(t)$ will look like a smooth ``S-curve'' as the price decreases, or as the
quality of the product increases. This is called the ``moving equilibrium'' model in the innovation diffusion literature \cite{Young}. 

The situation can change drastically when imitation is introduced. We keep here to a mean-field coupling to the average opinion, and comment on the
robustness of the results later, section \ref{sect3F}.  This global feedback effect simply shifts $F$ to $F+J m$, leading to a self-consistent equation:
\be \label{MF} 
m^* = 1 - 2 P_<(-F - J m^*).
\ee
The threshold model of Schelling and Granovetter \cite{Schelling, Granovetter} is a particular case, although usually expressed slightly differently. As above, 
call $N_+$ the number of agents deciding to attend a seminar, or join a riot, etc. Each agent has a preference expressed as a conformity threshold: 
if the observed number $N_+$ exceeds an $i$-dependent number that we set as $N(1-f_i)$, then agent $i$ joins as well the next time round.
 \footnote{In agreement with the above convention, $f_i$ large means that agent
$i$ is more prone to join, i.e. his threshold is lower.} This leads to a dynamical equation of the form:
\be
\phi_{t+1} = \int_{1-\phi_t}^\infty {\rm d}f \rho(f) = 1 - P_<(1-\phi_t),  \quad {\rm with} \quad \phi_t \equiv \frac{N_{+,t}}{N}.
\ee
The ``equilibrium'' value is given by the fixed point solution $\phi^* = 1- P_<(1-\phi^*)$ which coincides with Eq. (\ref{MF}) with $J=-F=1/2$. 
Eq. (\ref{MF}) contains $F$ as an extra degree of freedom, which allows one to treat both conformity and the moving equilibrium idea when $F$ is time dependent. Note that 
when all $J_{ij}$'s are positive (which is the case of the mean-field model), decision changes are `irreversible' when $F(t)$ has a monotone 
evolution in time. In other words, if $S_i$ flipped from $-1$ to $+1$ at a certain moment in history, it will never flip back to $-1$ later. 
This would not be true in the much richer (and more complicated) case where $J_{ij}$ can take both positive and negative signs. 

\subsection{Imitation vs. heterogeneity}
\label{sect3B}

Coming back to Eq. (\ref{MF}) when imitation is weak enough, one can expand the right hand side in powers of $J$, leading to first order to:
\be
m^* \approx \frac{m_0}{1 - 2 \rho(F) J}.
\ee
This equation shows that  the point where the slope of $m_0$ vs. $F$ is maximal
coincides (for symmetric distributions) with the point where the speed of variation of opinion changes is 
maximally amplified: imitation leads to exaggeration. As imitation becomes stronger, the above perturbative expansion breaks down. Still, the maximum slope of $m^*$ 
vs. $F$ increases and finally {\it diverges} for a critical value $J=J_c$, beyond which $m^*$ is a discontinuous function of $F$ (see Fig. 1, red line). Much as in the previous section, 
the self-consistent equation Eq. (\ref{MF}) has, for a range of $F$, three solutions for
$m$, one of which being unstable (see Fig. 1). In an economic context, it means that the demand for a product can be a multi-valued function of price,
i.e. the possible coexistence of a high demand and a low demand solution for the same price. This has been fully investigated in \cite{Nadal1,Nadal2,Nadal3}.

The solution chosen by the system depends on history. Suppose one starts with  $F = - \infty$ and $m = -1$ (zero demand), and slowly 
increases $F$ to make the alternative choice more appealing. The average opinion will first follow the lower branch until it jumps discontinuously to the upper branch, 
for a certain threshold field
$F_c(J)$ (and symmetrically on the way back, at $-F_c(J)$, as the field is decreased). Although Fig. 1 is qualitatively similar to the 
results of the previous section (which formally corresponds to $\sigma \equiv 0$, where $\sigma$ is the width of the distribution $\rho(f)$), there are many crucial differences. For example, the 
$m^*$ jump amplitude is not equal to $2$ but depends continuously on $J$ even when $\beta \to \infty$, and vanishes when $J \to J_c^+$ as $(J-J_c)^{\beta}$, where $\beta=1/2$
in mean-field or on tree-like graphs. Another difference is that the dynamics between $F = - \infty$ and $F_c(J)$ when the jump occurs is made of 
avalanches of different sizes, which we will discuss in section \ref{sect3D} below. A third difference is the existence of many more ``equilibrium states'', beyond 
the ones corresponding to the protocol that we describe here (i.e. $F = - \infty$ and all spins down at $t=0$). Other states can be 
reached, for example preparing the system with $m = 0, F=0$ one will follow a totally different hysteresis loop as $F$ is increased. 
This ``sensitivity to initial conditions'', or strong history dependence, makes the model quite appealing.  \footnote{These effects, strictly 
speaking, disappear in finite dimensional lattices because of nucleation. But the time scales associated with nucleation may be so large that
these metastable states still have a real existence.}

\begin{figure}
\begin{center}
\psfig{file=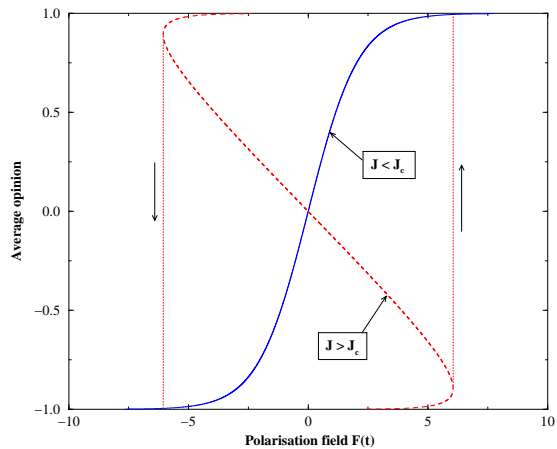,width=6cm,angle=270} 
\end{center}
\caption{Average opinion, or aggregate demand, or overall trust (etc.) $m$, as a function of the external
incentive field $F(t)$ in the {\sc rfim}. For an imitation parameter $J < J_c$, the curve is smooth. But when $J > J_c$ a
hysteresis effect appears with two possible stable states for the same value of $F$, plus one unstable state. As $F$ decreases the overall trust
remains high even for negative values of $F$, before suddenly collapsing at $-F_c(J)$ to the lower branch (as indicated by the arrow). The hysteresis effect means that
trust can only reappear if $F$ grows back beyond $+F_c(J)$, i.e.  much beyond the value where the crash happened.}
\label{Fig1}
\end{figure}

What sets the critical value $J_c$ of the interaction parameter beyond which strong herding effects occur? Dimensional analysis 
immediately implies that $J_c=A \sigma$ where $A$ a numerical constant (perhaps infinite) that depends on the detailed shape of
$\rho(f)$, and also on the topology of the graph. This means, as intuitively expected, that jumps and collective frenzy happen more easily in homogeneous populations (low $\sigma$)
than in strongly heterogeneous populations. Stability (of markets, democracies,...) requires diversity. For the mean-field model with a Gaussian $\rho$, $A = \sqrt{2\pi} \approx 2.5$. 
If the graph is a regular tree, then $A < \infty$ whenever the number of neighbours is $\geq 4$ \cite{Dhar}, but the transition disappears for a tree with 2 (i.e. a line) or 3 neighbours.
For these little connected lattices, interaction effects are never strong and there is no hysteresis phase. On a two-dimensional square lattice with each bond of strength $J$, the threshold is 
 $J_c \approx 1.56 \sigma$ \cite{Vives_num} and $J_c \approx 0.46 \sigma$ for a three-dimensional cubic lattice \cite{Sethna2} (both for a Gaussian $\rho(f)$).

\subsection{Critical scaling}
\label{sect3C}

The vicinity of the critical point $J \stackrel{<}{\sim} J_c$ reveals interesting critical properties, to a large extent
independent of the detailed form of $\rho(f)$. Noting $\varepsilon = J_c - J$ the distance from criticality,
one finds that the opinion slope ${\cal S}={\rm d}m/{\rm d}F$ takes a scaling form in a relatively wide region around of $F_c(J)$ \cite{Sethna2}:
\be \label{scaling}
{\cal S} = \frac{1}{\varepsilon}\,  {\cal G}
\left(\frac{F-F_c(J)}{\varepsilon^{3/2}}\right),
\ee
where the function ${\cal G}(x)$ can be computed explicitly \cite{Sethna_Dahmen,Michard}, and is universal. In particular, ${\cal G}(0)$ is a finite constant and ${\cal
G}(x \to \infty) \sim x^{-2/3}$. Eq. (\ref{scaling}) means that the slope, as a function of $F$, peaks at
a maximum of order $\varepsilon^{-1}$ and remains large on a small window of field of order $w \sim \varepsilon^{3/2}$. 
In other words, the height $h$ of the peak behaves as an anomalous power of the width $w^{-\kappa}$ with $\kappa=2/3$.
This is strikingly different from what is expected for any regular model, like the moving equilibrium model ($J = 0$), for 
which $h \sim 1/\sigma$ and $w \sim \sigma$, leading to $\kappa=1$.  Interestingly, this leads to testable predictions, which only rely on the assumption that the external incentive field
$F$ is a smooth function of time -- see section \ref{sect4A}. 

\subsection{Avalanches} 
\label{sect3D}

If now one `zooms' on finer details (on scale $1/N$) of the curve $m^*(F)$ in the vicinity of $J=J_c$, one 
discovers a very interesting structure. One finds that the 
evolution of $m^*$ is actually resolved in a succession of `avalanches' of
different sizes $s$, where the first opinion change induces (through interactions) a second one, then a 
third one, etc. After the $s$th one has flipped, the system is stable again, in the sense that for the current value of $F$ 
all $S_i$'s agree with the sign of the total incentives $U_i$.

The statistics of the avalanches is in fact easy to understand in mean-field, because the dynamics maps into the 
well known critical branching process. When one agent flips from $S_i=-1$ to $S_i=+1$, it induces a change of incentive 
$\Delta U = 2J/N$ for all other agents. For a given value of $F$ and $m^*(F)$, the agents that are on the verge of flipping have 
a preference $f_i$ precisely equal to $-F - Jm^*$. The average number of flips induced by the first one is thus given by \cite{Sethna_Dahmen}:
\be\label{n-def}
n = N \rho(-F -Jm^*) \times \frac{2J}{N}.
\ee
Each of the newly converted agents can convert on average $n$ new ones, etc. In mean-field, this is equivalent to the Galton-Watson
branching process with branching ratio $n$ \cite{GW}. It is well known that when $n < 1$, avalanches are all of finite size, whereas when $n > 1$
some avalanches will diverge and sweep the whole system. In the region of $n \stackrel{<}{\sim} 1$, the distribution of avalanche sizes is a truncated power-law:
\be\label{avalanches}
P(s) \propto_{s \gg 1} s^{-3/2} e^{-(1-n)^2 s}.
\ee
Now, let us explain why $n=1$ corresponds to the critical point $J=J_c$, $F=F_c$. Eq. (\ref{MF}) can be written $m^* = G(m^*)$, with $G(m)=1 - 2 P_<(-F - J m)$. 
One thus sees that the function $G(m)$ must cut the first diagonal once for $J < J_c$ and three times for $J > J_c$. The point $J=J_c$ 
must correspond to the case where $G(m)$ is {\it tangent} to the first diagonal at the point of crossing. Therefore, at the critical point,
\be
G'(m^*) = 2J \rho(-F -Jm^*) = 1,
\ee
which is indeed equivalent to $n=1$ (see Eq. (\ref{n-def})). Interestingly, exactly the same scenario occurs in the so-called Fiber-Bundle model for fracture \cite{Hansen_fiber,Rava}, which is very close in spirit
to the above models. In a nut-shell, the Fiber-Bundle model assumes initially $N$ intact fibers in parallel, subject to an external force $NF$ that is equally distributed over all fibers. 
Fibers are heterogeneous, and when the force acting on the $i$th one exceeds a certain threshold $\theta_i$, the fiber breaks and the external force is redistributed over all 
remaining fibers. It is clear that the fraction $\phi$ of intact fibers obeys the following self consistent equation:
\be
\phi^* = \int_{F/\phi^*}^\infty {\rm d}\theta \rho(\theta) = P_>\left(\frac{F}{\phi^*}\right),
\ee
where $\rho(\theta)$ is the distribution of fiber strengths. For $F < F_c$, the above equation has generically two solutions that merge for $F=F_c$ and disappear for $F > F_c$ (i.e. the
bundle breaks apart). Again, when $F=F_c$ the function $P_>(\frac{F_c}{\phi})$ has a slope equal to one at $\phi^*$. The same analysis as above can be carried over for the avalanches in 
this model as well. \footnote{The exponent $-3/2$ in the avalanche size distribution is the same as the one appearing in the first return time distribution of a one-dimensional random walk. 
Indeed, the critical branching process can be mapped onto that problem, see \cite{Sornette90s,Hansen_fiber} for more elaborations on this point.}

This analysis offers an enticing microscopic picture of how large opinion swings actually develop in a population: as a branching process that generates 
avalanches. Most of them are small, but some may involve an extremely large number of individuals, without any particularly large change of external conditions
(i.e. change in $F(t)$). The power-law distribution that appears is tantalizing and might
enable one to understand why bursts of activity and power-laws appear in the distribution of returns in financial markets, for example. But there is at present no 
convincing argument that would make this analogy sharp. 

\subsection{Delays and trends} 
\label{sect3E}

The above analysis assumes that the external incentive field $F(t)$ varies extremely slowly, so that the state of the system can follow adiabatically the equilibrium 
condition Eq. (\ref{MF}). It may be useful to generalize the model to include lag effects, and to write the following time evolution equation \cite{Young}:
\be\label{MFdyn}
\frac{dm^*}{dt} = \mu \left[1 - 2 P_<(-F(t) - J m^*) - m^*\right],
\ee
where $\mu^{-1}$ defines an equilibration time scale, i.e. the time needed for agents to adapt to the new piece of information that changed $F(t)$. If $\mu^{-1}$ is very short 
compared to the evolution time of $F(t)$, we are back to the previous analysis. Otherwise, lag effects matter  and could for example affect the avalanche size distribution $P(s)$  
for large $s$. 

Another potentially interesting generalisation is to include the influence of past trends. People are not only sensitive to the contemporaneous behaviour of others, they also take cues from
past patterns. An obvious candidate is the speed at which $m^*$ has varied in the past, which may give a sense of urgency and adds to the external incentive $F(t)$. The simplest model that 
encodes this would be Eq. (\ref{MFdyn}) with $F(t)$ replaced by:
\be
F(t) \longrightarrow F(t) + g  \int_{0}^t {\rm d}t'  e^{-\Gamma(t-t')} \frac{dm^*}{dt'},
\ee
where $\Gamma^{-1}$ is the time scale over which the past trend is computed and $g$ the strength of the coupling to past trends. This could be an interesting model for 
the dynamics of financial markets, similar to the one discussed in section \ref{sect6C}. Other examples of feedback between past history and present decisions will also be reviewed there.

\subsection{Phase diagram for noisy decisions} 
\label{sect3F}

Before moving to more quantitative evidence for the type of effects that the {\sc rfim}  is supposed to capture, it is important to end the present section with a discussion of the role of the 
coefficient $\beta \equiv 1/T$ in the decision rules. Up to now, we considered that $\beta = \infty$ -- i.e. decisions are `rational' and conform to the overall incentive, $S_i = \mbox{sign} [U_i]$.
Introducing some amount of noise does not immediately destroy the above picture: for a given imitation strength $J$, there is still a critical value $\sigma_c(T)$ of the heterogeneity below
which jumps occur and above which the demand curve becomes smooth. Similarly, for a fixed $\sigma$, jumps remain until decision noise becomes larger than a $\sigma$-dependent value $T_c(\sigma)$ 
(i.e. for $\beta$ smaller than $\beta_c=1/T_c$). The situation is summarized in Fig. 2, which shows the region of the $(T,\sigma)$ plane where jumps and hysteresis occur, located below the red line. 
Above that line (i.e. where heterogeneities and/or noise are strong enough, there is a single equilibrium and a unique demand/opinion curve). The model in section \ref{sect2A} corresponds to the $x$-axis 
of this graph ($\sigma=0$), whereas the results of the present section correspond to the $y$-axis ($T=0$). From the point of view of potential applications of the model, it is reassuring to 
see that the whole scenario predicted by the {\sc rfim}, and in particular the transition between a ``crisis'' phase and a smooth phase, is quite robust against details (noise, network topology, 
shape of the probability distribution of preferences, etc.). \footnote{This universality was also noted and emphasized in  \cite{Nadal2,Nadal3}.} Finally, note that there is a second, green line in 
Fig. 2 \cite{Krzakala}, separating a regime of strong history dependence where the number of equilibrium states is exponential in the number of agents and hysteresis sub-loops are stable, 
from a phase that is more similar to what happens in the homogeneous case $\sigma=0$, with a single (large) hysteresis loop. \footnote{For other recent examples of agent based models with an exponential number of 
equilibria, see \cite{Farmer_new,Lucas}.}

\begin{figure}
\begin{center}
\psfig{file=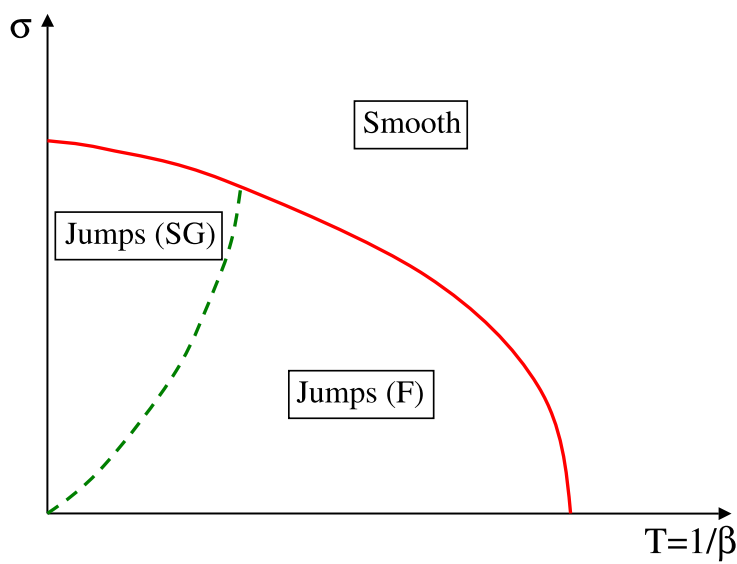,width=7cm} 
\end{center}
\caption{Qualitative phase diagram, in the plane $(T=1/\beta,\sigma)$, of the mean-field {\sc rfim}. The model in section \ref{sect2A} corresponds to the $x$-axis 
($\sigma=0$), whereas $T=0$ {\sc rfim}  of the present section corresponds to the $y$-axis. The aggregate behaviour is smooth above the red boundary, but 
exhibits jumps and hysteresis below. An even more complex Spin-Glass-like phase, with an exponentially large number of metastable states, appears to the left of the 
dotted green line (see \cite{Krzakala} for more details).}
\label{Fig2}
\end{figure}

\subsection{Summary}
\label{sect3G}

In summary, the transposition of the {\sc rfim} to opinion/demand formation predicts that the average opinion evolves,
as a function of the global solicitation, very differently if imitation
effects are weak (in which case the evolution is smooth) or if imitation exceeds a certain threshold, in
which case aggregate quantities are discontinuous and catastrophic avalanches may be triggered. More
quantitatively, the model predicts that in generic situations, the speed of change should peak more and more as the
critical point is approached. As a function of time, the speed of change exhibits a maximum of height $h$ and of 
width $w$, the two being related by $h \sim w^{-2/3}$. This is very different from non-critical adoption models, such 
as the ``moving equilibrium'' model, for which $h \sim w^{-1}$. These predictions seem to be qualitatively and even 
quantitatively relevant in several instances that we describe in the next section.

\section{The {\sc rfim}  scenario: empirical evidence and possible applications}

It is now a rather well established fact that people do not make decision in isolation but rely on the choice/opinion of others,
see \cite{Raafat,Baddeley} for a detailed discussion and many references. Recent work includes in particular the artificial culture market of Salganik et al. \cite{Salganik}, 
the analysis of earning forecasts by expert analysts \cite{Guedj}, or the wisdom of crowds \cite{Helbing_crowds}. 
Of course, anecdotal evidence concerning famous financial bubbles (like the tulip-mania in Amsterdam and other examples of `madness of crowds'
\cite{Thistime}) or famous fads \& fashions (like Louis XIV's wig, Catherine de Medicis' forks and Richelieu's rounded knives, etc.) make 
it quite obvious that social pressure can be disturbingly dominant. Keynes conceived imitation (herding) as a response to uncertainty and
to the recognition of one's own ignorance: people follow the crowd because they think (or fear) that the rest of the crowd might be better informed. \footnote{
Peyton Young in \cite{Durlauf_Book} contrasts instrumental conformism, when it is beneficial to do what others do, to informational conformism, when we conform to what others do 
because it conveys information on best actions. The problem, we believe, is that often nobody really has any idea of what's going on, so the ``information'' provided by
the others actions is close to zero.} 
Gigerenzer provides a complementary interpretation \cite{Gigerenzer,Gigerenzer_Book}: people tend to follow rules of thumb and routines. Imitation qualifies well as a `fast and frugal' 
heuristic, which furthermore has minimal information costs. However, a recurrent problem has been to disentangle effects induced by direct 
imitations or social pressure from correlations that would appear just because people have the same information or the same cultural bias, etc. \cite{Manski}

\subsection{Some quantitive evidence}
\label{sect4A}

In that respect, the robust prediction of the {\sc rfim}  alluded to in the previous section allows one to test for the existence of herding effects 
with minimal extra assumptions. This is the program that we followed with Q. Michard in \cite{Michard}. The idea was to compare the dynamics 
of the very same phenomenon (for example, the adoption of cell phones) in different countries, supposedly characterized by different imitation 
parameters, and different distributions of willingness to adopt. Still, assuming (a) that the speed of evolution of the external incentive field
$F(t)$ is similar in all countries, and (b) that the ratio $J/\sigma$ is not too far from the critical value of the model  \footnote{The {\sc rfim}  is 
actually well known to have a rather wide critical region, as emphasized in \cite{Sethna2}.} then the scaling prediction Eq. (\ref{scaling}) should hold, leading 
to a universal relation between the height $h$ and the width $w$ of the evolution speed ${\cal S}={\rm d}m/{\rm d}F = v^{-1} {\rm d}m/{\rm d}t$,
where $v = {\rm d}F/{\rm d}t$ is the speed of evolution of the external incentive field, assumed to be independent of the country or population.
If the distance from the critical point varies across the set of available countries or crowds, ${\cal S}$ should reveal peaks with different heights and widths, 
but all related by $h \sim w^{-\kappa}$ with $\kappa=2/3$ if imitation is strong. \footnote{We emphasize again here that if $J=0$, or if the 
variations were due to different speeds $v$ in different countries, one should observe $\kappa=1$. The way $h$ and $w$ are extracted from data is detailed in 
\cite{Michard}.} We gathered data concerning (a) the drop of
birth rates in European countries in the second half of the XXth century \cite{Fertility} (b) the increase of cell phones in Europe in the 90's, (c) the way clapping dies out at the end 
of music concerts and (d) crime statistics in different US states in the period 1960-2000. Although social influence have been argued to
be important in the latter case too \cite{JS}, our data set did not show enough idiosyncratic variations across states to be exploited. In the first three cases, 
we find that our data fits well the picture suggested by the model, and that collective effects seem to be present, with indeed $\kappa \approx 2/3$ -- see Fig. 3. 
The example of clapping is interesting because it is very close to being a controlled experiment \cite{Michard}. 
In that case, we observed both continuous and abrupt endings, as predicted by the model. 
Challet et al. \cite{Unilever} analyzed data from a Swiss on-line supermarket, and report $\kappa = 0.55 \pm 0.16$ for soap consumption. More data would be 
needed to ascertain the relevance of the {\sc rfim}  scenario in real situations. In particular a finer time resolution would allow one to test one of the 
crucial predictions of the model: the presence of bursts, or avalanches, with a truncated power-law distribution of sizes (see Eq. (\ref{avalanches})). This would be a clear-cut
signature of imitation cascades. \footnote{see also \cite{Cascades,Cascades2} for experiments on information cascades and \cite{tip} for some empirical evidence of tipping 
points in urban segregation phenomena. We also refer the reader to \cite{Blume} for a review on ``econometric'' approaches to social phenomena.}

\begin{figure}
\begin{center}
\psfig{file=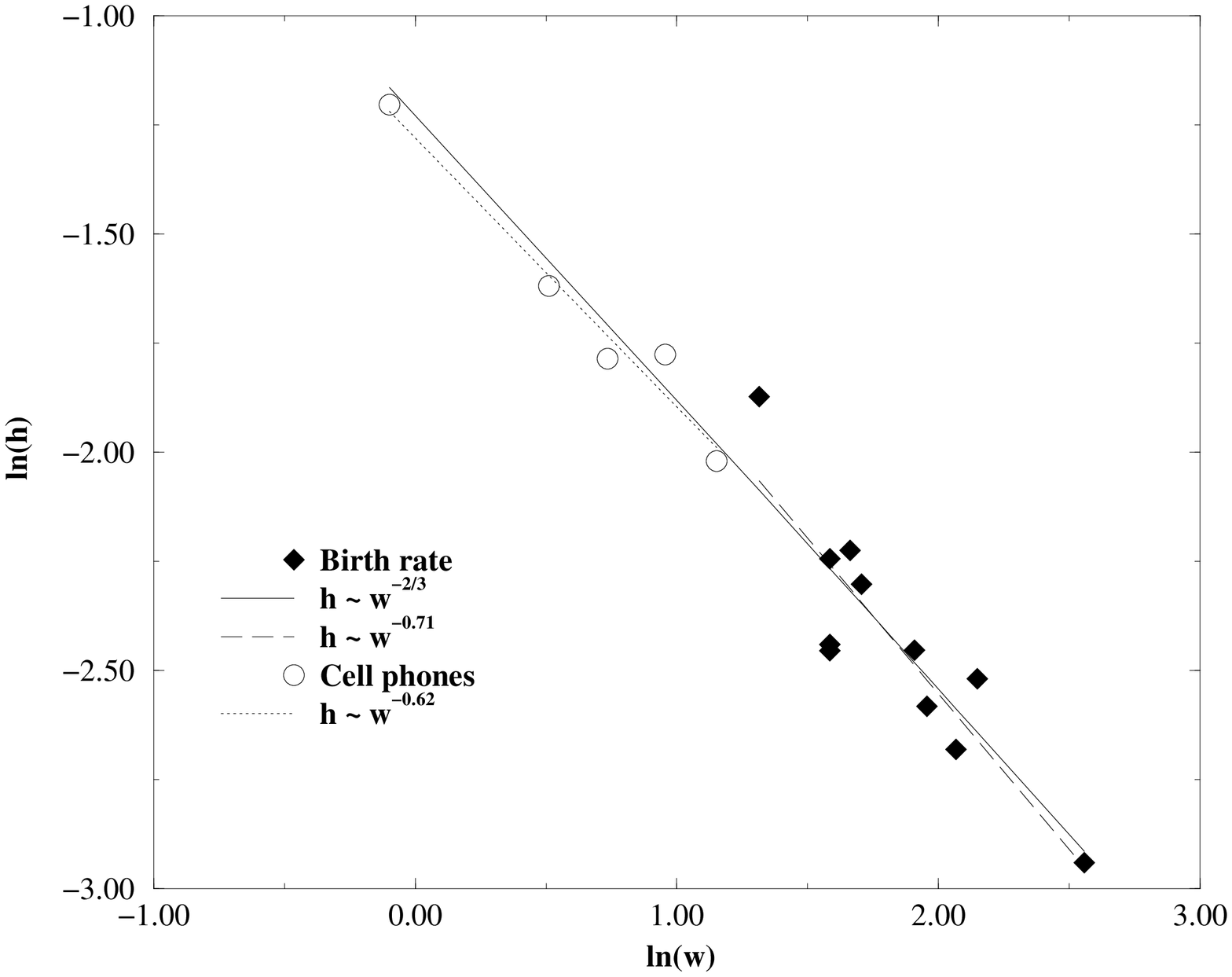,width=5cm,angle=270} \psfig{file=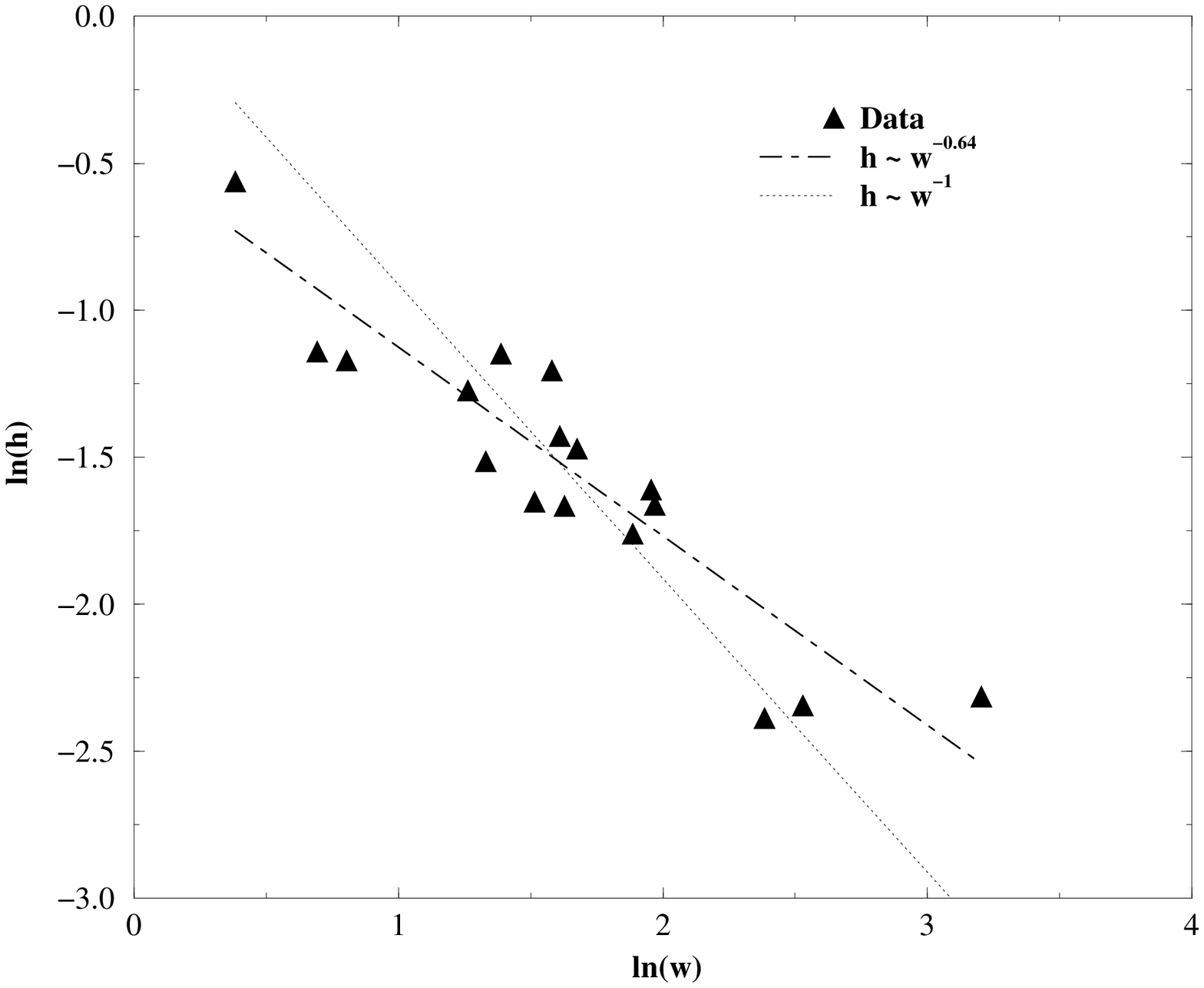,width=5cm,angle=270}
\end{center}
\caption{Left: Height of the peak $h$, vs. width of the peak $w$, in a log-log
scale, both for birth rates and for cell phones, for different European countries (see \cite{Michard}). A best fit leads to a slope (in log scale) of, respectively: $-0.71 \pm 0.11$
and $-0.62 \pm 0.07$. The mean field  {\sc rfim} critical prediction $h \sim w^{-2/3}$ is shown for comparison. Right: Height of the peak $h$, vs. width of the peak $w$, in a log-log
scale, for 17 applause endings. A best fit leads to a slope of $-0.64 \pm 0.07$, again very close to the mean-field prediction of $-2/3$. We also show for comparison the $-1$ slope 
corresponding to a `trivial' behaviour, $h \sim 1/w$ (thin dotted line).}
\label{Fig3}
\end{figure}

\subsection{Metaphors}
\label{sect4B}

In any event, the model has a huge narrative potential, and is well adapted to describe, qualitatively or quantitatively, many situations. For example, 
Nadal et al. \cite{Nadal1,Nadal2,Nadal3} have analyzed the model in great detail in the case of a product sold at price $p$ by a monopolist supplier, and in the presence of 
imitation effects on the demand side. They note that the monopolist has two different strategies: either to attract a large fraction of buyers at
low prices or to target few buyers willing to pay high prices. There is again a first order transition where the optimal strategy switches abruptly from one to the other \cite{Nadal3}. 
They further note that, interestingly, {\it the optimal price is generically just slightly
below the price at which the high demand equilibrium disappears. A small
change in the customers characteristics may lead to a decrease of this critical
price. If this change is not anticipated by the seller, the posted price may
become greater than the critical price, resulting in a collapse of the demand}, and they relate this observation to Becker's interpretation of 
the ``crowded restaurant paradox'' \cite{Becker_restaurant}. 

The sudden shift from the high demand branch to the low demand branch is also a natural 
mechanism for crashes in financial markets. We believe \cite{Risk}, like many others, in the critical importance of trust as a determining factor for the prosperity of economies, 
and for securing well-functioning 
financial markets. Trust is a collective asset that allows efficient coordination and tremendously accelerates business. The most illuminating 
example is fiat money, that can only exist if people trust that a piece of paper will not be worthless tomorrow -- and of course if everybody 
gives value to a dollar bill, the dollar bill is valuable. Clearly, money is more efficient than barter because it solves the otherwise
daunting problem of matching individual preferences. On the other hand, because trust is an immaterial sentiment, it can evaporate overnight. 
Changing one's mind can be instantaneous, the speed of change is not limited by physical constraints. This 
is one of the reason why crisis, crashes and bank runs can occur (this point of view is forcefully expressed in, e.g. \cite{Hosking,Marsili_trust} and refs. therein). 
Think for example of the sudden freezing of the money markets just after Lehman's default: clearly, banks stopped lending {\it because} all other banks stopped lending. 
Credit means: ``I believe ({\it credo}) that I will see my money back some day'', and this belief can disappear even if no destruction of wealth has physically taken place. 
Established trust is a situation where the subjective probability of default or of failure is very small. 

Several quantitative models for trust collapse have been studied in the past few years, see e.g \cite{Marsili_trust,Marsili_Kirman,Lorenz,Holyst} and 
refs. therein. Some are directly related to the {\sc rfim}, which can indeed be formulated as a model for trust cycles, associated to bubbles and busts, 
bank runs and hyper-inflation. Suppose that for some reason the world is in a euphoric state, corresponding to $F \to +\infty$ and $m=1$ -- everybody trusts the system. 
Fiat money is considered as absolutely safe. 
Now, imagine that for some reason the objective trustworthiness of the system $F$ decreases. This could be due to a rise of inflation, or to over-valued 
prices of assets or houses that appear far above any reasonable fundamental value. This could also be due to political instability, or an increase in 
any measured or perceived risk. The most risk adverse individuals will act first and sell their assets, or take their money away from the bank, whereas most people 
firmly believe that the system is safe. \footnote{I personally know people who cashed their money just after Lehman's default, and left
their bank with a plastic bag full of bank notes.} Still, these initial ``irrational'' reactions seed wariness in the minds of others -- what if they were right? 
Maybe I should do the same? As dark clouds accumulate, it is easy to guess what will happen. Either imitation effects are small 
and people sell one by one their assets, leading to a slow decay of the stock markets or a soft landing of the business cycle. 
Or imitation is beyond the critical threshold and after an extended period where confidence is only maintained because everybody else is confident 
(as in a bubble), a small exogenous event (i.e. a small further decrease of $F$) may be enough to trigger global panic, i.e. crashes, bank runs, hyper-inflation, etc. 
Because the system is hysteretic, the fundamental information $F$ must grow back much beyond the point where the crisis occurred for 
people to regain confidence (see Fig. 1).

One can think about the model slightly differently. Imagine that the external world, captured by $F$, is fixed, but that the imitation parameter $J$
can vary in time. This is quite realistic: in periods when uncertainty is small, people can rely on fundamental information to make a judgement. 
When uncertainty increases, it becomes more and more tempting to take one's cue from the crowd. As $J$ increases past $J_c$, an initially weakly 
polarized state $m \sim 0$ can suddenly evolve towards $m = \pm 1$ (see Fig. 1). This is a kind of volatility feedback effect, where past volatility may trigger
more herding effects, finally leading to a crash (see e.g. \cite{Cont,Curty,Harras,Harras2}).

\subsection{Spatial effects}
\label{sect4C}

Another interesting topic is the spatial structure of the incentive field $F$, which can be interpreted in the {\sc rfim}  setting as the 
average willingness (to pay/adopt/trust, etc..), around which an idiosyncratic component $f_i$ is added. It might be expected that on several issues, the 
average willingness strongly depends on regional characteristics and culture. One specific example for which spatially resolved data is available are 
elections. The simplest observable is the turnout rate, which reveals persistent patterns, both in time and in space -- see \cite{Borghesi_election1,
Borghesi_election2} and Fig. 4. Calling $\phi(\vec r)$ the turnout rate on a given election in a city located at $\vec r$, one can define the logarithmic 
turnout rate $\tau(\vec r)$ as:
\be
\tau = \ln\left(\frac{\phi}{1 - \phi}\right), \qquad \tau \in (-\infty,+\infty)
\ee
The pattern of Fig. 4 suggests a non-trivial bahaviour of the spatial correlation of the $\tau$ field, which is found to decay roughly logarithmically with distance:
 \footnote{In the following expression, $\langle ... \rangle$ denotes an averaging over all positions $\vec R$.}
\be\label{log}
C(\vec r) = \langle \tau(\vec R) \tau(\vec R + \vec r) \rangle - \langle \tau(\vec R) \rangle^2 \approx A - B \ln r, \qquad r = |\vec r|,
\ee
as shown in Fig. 4, inset. This logarithmic dependence holds well for different elections and different countries, up to distances $r$ comparable to the size of the 
country \cite{Borghesi_election2}. Another striking observation is that the voting habits of the different regions seem to be extremely persistent historically \cite{Borghesi_election1}.

Assuming a logistic distribution for the idiosyncratic willingness to vote, one finds: \footnote{Note that here we neglect the immediate influence of others 
in the actual decision to vote, i.e. the $J S_i S_j$ interaction term. See the discussion of this point in \cite{Borghesi_election1}.} 
\be
\phi = \frac{1}{1 + e^{\beta F}} \longrightarrow \tau = \beta F,
\ee
where $F$ is the (local) average willingness to vote. From the above identification, we conclude that this average willingness itself has logarithmic 
correlations in space and long-range correlations in time.  This suggests the existence of a ``cultural field'' that keeps the memory independently of the presence
of particular individuals.

A simple and rather natural idea is to decompose the time dependent incentive field $F(\vec r,t)$ as a sum of a $\vec r$-independent component $\overline{F}(t)$ (which measures the average interest in the election), 
a slowly varying ``cultural'' field $\psi(\vec r,t)$ that we want to model, and a white noise (in space) contribution $\epsilon(\vec r,t)$, which models city-specific idiosyncraties and only has
short-ranged spatial correlations \cite{Borghesi_election1,
Borghesi_election2}. The cultural field $\psi(\vec r,t)$ should evolve due to (a) cultural ``shocks'' that contribute to changing the overall mood in a given city (such as the 
closing down of a factory or of a military base, important changes in population, etc.) and (b) diffusion, as a consequence of the exchange of ideas and opinions between nearby cities.
Through human interactions, the cultural differences between nearby cities tend to narrow, leading to a Laplacian term in the evolution equation (in the continuum limit). We therefore 
proposed to write the following noisy diffusion equation for $\psi(\vec r,t)$ \cite{Borghesi_election1,Borghesi_election2}:
\be\label{diff}
\frac{\partial \psi(\vec r,t)}{\partial t} = D \nabla^2 \psi(\vec r,t) + \eta(\vec r,t),
\ee
where $D$ measures the speed of diffusion of the cultural field \footnote{Taking the relevant inter-town distance to be $\sim 10$ km and the time for opinions to get closer to 
be a few months, one can estimate $D$ to be of the order of magnitude of a few hundreds km$^2$ per year.} and $\eta(\vec r,t)$ a white noise field. If cities were located on the nodes of a regular grid, 
it would be easy to show analytically that the stationary correlation function of the field $\psi(\vec r,t)$ decays as a logarithm of distance, as given by Eq. (\ref{log}). 
However, the spatial distribution of cities in countries 
is in reality quite strongly heterogeneous, which leads to significant concavity when $C(r)$ is plotted as a function of $\ln r$ \cite{Borghesi_election2}. Quite remarkably, this concavity is very similar to what is observed for the empirical correlations 
\cite{Borghesi_election1,Borghesi_election2}, which tends to confirm that the long-range spatial correlations observed in election turnout rates can indeed be traced to the presence of an underlying cultural field 
with a diffusive dynamics.

\begin{figure}
\begin{center}
\psfig{file=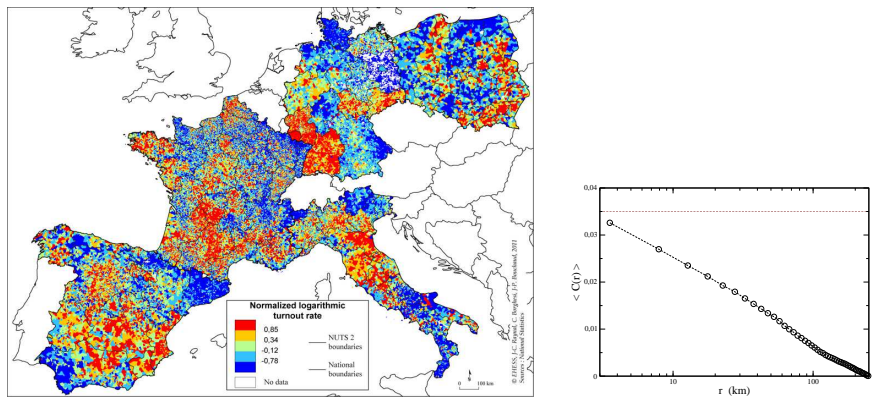,width=7cm}\hskip 0.5cm \psfig{file=cov-tau-moy-fr.eps,width=4cm}
\end{center}
\caption{Heat map of the normalized logarithmic turnout rate $\tau(\vec r)$ for the 2004 European Parliament election in France, Germany, Italy, Poland and Spain. Note the strong, long ranged correlations in the
patterns. Inset: Average of spatial correlations $C(r)$ over 20 different French elections, in a semi-logarithmic plot to reveal the quality of the logarithmic behavior, Eq. (\ref{log}). See \cite{Borghesi_election2}
for details.}
\label{Fig4}
\end{figure}

While preparing this review, we discovered that a diffusion equation very similar to Eq. (\ref{diff}) was in fact proposed by Schweitzer and Hoylst in 2000 \cite{Schweitzer,Schweitzer2} to model opinion 
formation. They call $\psi$ the ``information field'' and their equation can be written, in the present context, as:
\be
\frac{\partial \psi(\vec r,t)}{\partial t} = D \nabla^2 \psi(\vec r,t) + k [\phi(\vec r,t) - \psi(\vec r,t)].
\ee
The last term describes a coupling between the past realized turnout rates and the cultural field, i.e. large participation rates tend to motivate people to vote more the next time round. 
However, the presence of such a  term in the diffusion equation would cut-off the logarithmic dependence of $C(r)$ beyond a distance $\sqrt{D/k}$. Since this is not observed empirically, 
$1/k$ is probably large (20 years or more) -- at least in the context of election participation. 

The idea of a persistent cultural (or information) field that evolves according to a noisy diffusion equation seems quite important. If true, many other social phenomena like consumption habits, behavioral biases, etc.
should exhibit long ranged (logarithmic) spatial correlations. More empirical work on the spatial correlations of these decisions, for example based on consumption data, in different situations and in different countries,
would be very valuable -- although the results on election turnout in many countries are already quite encouraging \cite{Borghesi_election2}.

\subsection{Multiple choices} 
\label{sect4D}

A generalisation of the {\sc rfim}  setting to multiple choices is interesting for obvious reasons: when in a bookstore, supermarket or 
record shop, the choice is not between two products but between thousands, or even millions of products. The same is true of the stock market -- 
one can choose to invest in thousands of different stocks. Again, the choice of others is an important piece of information that we take into account before buying 
a book or deciding on a movie. It is plausible that herding effects play a role in the appearance of Pareto-tails in the measure of success 
(book sales, movie attendance, smart phones, etc.). A natural generalization of Eq. (\ref{rfim}) to an M-choice situation reads \cite{Borghesi_choice}:
\be 
n_i^\alpha(t+1) = \Theta(F^\alpha + f_i^\alpha + JM \pi_\alpha(t)),
\ee
where $n_i^\alpha$ is the consumption of product $\alpha=1, \dots, M$ by agent $i=1,\dots, N$, $\Theta(u > 0)=1$ and $\Theta(u \leq 0) = 0$, and $\pi_\alpha$ is the relative consumption of product $\alpha$, i.e. 
\be
\pi_\alpha = \frac{\sum_i n_i^\alpha}{\sum_{j,\beta} n_j^\beta}.
\ee
The field $F^\alpha$ describes the intrinsic attractivity of product $\alpha$, while $f_i^\alpha$ is the willingness to pay of agent $i$, to acquire 
product $\alpha$. As above, the $J$ coefficient measures the strength of social influence, and a factor $M$ is added to make sure that social 
influence remains of order unity when $M \to \infty$. 

The most interesting question about such a model is to know whether the realized consumption is faithful, i.e. whether or not the actual choice of the different items 
reflects the `true' preferences of individual agents, as would be the case in the absence of interactions ($J = 0$). Based
on the phenomenology of the {\sc rfim}, one expects that this will not be the case when $J$ is sufficiently large. The exact solution of the model 
in the limit $M \to \infty$ indeed shows that a phase transition takes place for $J = J_c$, where $J_c$ can be exactly computed in terms of the 
distribution of the $F$s and the $f$s, assumed to be IID random variables. For $J > J_c$,  strong distortions
occur, meaning that the realized consumption will (i) ``condense'' into a small subset of all possible choices, (ii) violate the natural ordering of individual preferences
and (iii) become history dependent: a particular initial condition determines the 'winners' in an irreproducible and
unpredictable way \cite{Marsili_Raf1}. Instead of all products getting a $\sim 1/M$ share of the market, some (but not necessarily the best ones, with large $F$s) become `hits' and attract a disproportionate 
number of buyers. This is a possible mechanism for the appearance of Pareto tails in most measures of success \cite{Zajdenweber,Redner}.
Similar effects were observed in the artificial cultural market of Salganik et al. \cite{Salganik}, which allowed them to conclude that increasing the strength of 
social influence increases both the inequality of market shares and the unpredictability of success (i.e. which 
products become winners).

The model can be transposed to other interesting situations, for example that of industrial production, for which
one expects a transition between an archaic economy dominated by very few products and a fully diversified economy,
as the dispersion of individual needs becomes larger as time passes.

\section{Choice theory and decision rules}

\subsection{The logit rule: arguments and extensions}
\label{sect5A}

The literature on choice theory is extremely vast and has been beautifully reviewed in \cite{Anderson_Palma} (although the field has progressed since 1992). We want here to restrict to a few comments that 
may help to understand some of the problems at stake. In a nutshell, classical choice theory assumes that the probability $P_\alpha$ to choose alternative $\alpha$ in a set 
${\cal A}$ of different possibilities can be written as:
\be\label{logit3}
P_\alpha = \frac{e^{\beta U_\alpha}}{\sum_{\gamma \in {\cal A}} e^{\beta U_\gamma}},
\ee
where $U_\alpha$ is the ``utility'' of alternative $\alpha$ and $\beta$ the ``noise'' parameter that allows one to interpolate between equiprobable choices ($\beta=0$) and
utility maximisation ($\beta \to \infty$). Before discussing the arguments that have been put forth to justify the above ``logit rule'', one should specify what one really 
means by  ``the probability $P_\alpha$ to choose alternative $\alpha$''. Two quite different interpretations can be found in the literature (see \cite{Anderson_Palma}, pp. 30-33):
\begin{itemize}

\item (1) One is that each agent, as a function of time, flips between different alternatives. At each time step the probability to choose $\alpha \in {\cal A}$ is the above $P_\alpha$. 
This could be due to an imperfect knowledge of the utilities $U_\alpha$, or to ``irrational'' effects such as illusions and intuitions, cognitive limitations, etc. It could also
be due to a truly time evolving context or environment, not described in the theory, but that randomly affects the relative utilities of the alternatives, in such a way that 
the {\it perceived} utility of agent $i$ at time $t$ is $\hat U^i_\alpha = U_\alpha + \epsilon^i_\alpha(t)$, where $\epsilon^i_\alpha(t)$ are time dependent random variables. 
In this interpretation, $P_\alpha$ makes sense even for a single agent: it is the probability that, at a given instant of time he or she chooses $\alpha$.

\item (2) The second interpretation assumes that any given agent $i$ always makes the same choice, based on the maximisation of perceived utility $\hat U^i_\alpha = U_\alpha + \epsilon^i_\alpha$,
where $\epsilon^i_\alpha$ are again random variables but now fixed in time. In this case, $P_\alpha$ is the probability that a given agent in the population has forever chosen $\alpha$.

\end{itemize}

At the aggregate level, the two interpretations may look indistinguishable (see \cite{McFadden}), and this is indeed true for non-interacting (independent) agents. In the presence of interactions, 
however, the two formulations lead to different results for the aggregate behaviour (see section \ref{sect5B}). The first interpretation corresponds to the $x$-axis in the phase diagram of Fig.2, 
while the second corresponds to the $y$-axis. 

The proposed justifications for the exponential form of the probabilities in Eq. (\ref{logit3}) are of three types \cite{Anderson_Palma}: (a) axiomatic  \cite{Anderson_Palma}; (b) based on specific assumptions about the
random variables $\epsilon^i_\alpha$ that affect the perceived utilities; (c) based on entropy arguments.  On point (b): if the $\epsilon$'s are IID random variables with a Gumbel (double-exponential) 
distribution, then it is possible to show that the probability is indeed given by Eq. (\ref{logit3}) \cite{McFadden}. However, the deep reason for choosing a Gumbel distribution has remained elusive -- the fact 
that the Gumbel distribution is max-stable does not seem to help (see however \cite{MG}, pp 32-33). On point (c): one can argue that agents do not want to maximize the expected utility alone, but take into account the information 
cost (or the entropy) associated to the weights $P_\alpha$. As is well known, the maximisation of $\beta \sum_\alpha P_\alpha U_\alpha -  \sum_\alpha P_\alpha \ln P_\alpha$ indeed directly 
yields Eq. (\ref{logit3}). This has been called the {\it variety seeking} behaviour of agents in \cite{Anderson_Palma} (pp. 78-79), and the exploration-exploitation compromise in \cite{Nadal_Kirman}, which might account for some inherent uncertainty about the time stability of utilities (see also \cite{Marsili_Logit}). 
Still, all these arguments look a bit weak; \footnote{Other arguments, inspired by statistical physics, could go as follows \cite{Marsili_Logit}: imagine a choice consisting in two sub-choices concerning issues in disjoint sets  ${\cal A},{\cal B}$. These two 
sub-choices are furthermore assumed to be independent, i.e. the choice of an alternative $\alpha$ in ${\cal A}$ does not impact the utility of any of the alternatives $b$ in ${\cal B}$, and vice-versa. 
The utilities are therefore additive in that case: $U_{\alpha \oplus b} = U_\alpha + U_b$. Since the sub-choices are independent, one should also have $P_{\alpha \oplus b} = P_\alpha \times P_b$. Looking for probabilities
that depend on the total utility of the choice therefore selects the exponential form. One could also try to replicate the usual canonical construction of the Boltzmann weight, by arguing that a 
choice is never in isolation but interacts with many other choices, while the agent is only interested in the total utility. Sub-optimal choices (corresponding to a finite $\beta$) are allowed because
they only have a small contribution to the total utility. However, these arguments are somewhat {\it ad-hoc}.} 
the ultimate justification seems to be that the logit rule is mathematically very convenient \cite{Anderson_Palma}, and allows one to use well known results 
in statistical physics because the logit rule is equivalent to the Boltzmann-Gibbs measure. 

We will explore more the dynamical interpretation (1) above, which seems to us more realistic. People can change their minds and not make the same decision all the time, even when confronted to 
the very same information (see the discussion in \cite{Anderson_Palma}). The idea of introducing a
``tremor'' noise in the utility dates back to Thurstone \cite{Thurstone}. If there is time correlations in the noise, and no learning or feedback from the past that induce some systematic evolution 
of the utilities, one can think of a dynamical model of decisions as the following Markov chain. At each time step, agent $i$, having made choice $\alpha$, reviews his alternatives with probability 
$\mu$ and does not bother with probability $1 - \mu$. The possible alternatives $\gamma \neq \alpha$ are in a set ${\cal A}_\alpha \subseteq {\cal A}$. Often people do not even consider all 
alternatives simultaneously but sequentially, so just pick one possible $\gamma$, and decide whether it is worth going for that one. (We make this assumption to simplify the following discussion.) 
It is reasonable to postulate that the probability for this alternative $\gamma$ to be adopted is equal to the probability that the noisy utility contributions are such that:
\be
\epsilon^i_\gamma(t) - \epsilon^i_\alpha(t) > U_\alpha - U_\gamma.
\ee
Alternative $\gamma$ is therefore adopted with probability $Q_>(U_\alpha - U_\gamma)$, where $Q_>$ is the complementary cumulative distribution function of the noise 
differences. The time-dependent probabilities $P_\alpha(t)$ therefore obey the following Master equation:
\be
P_\alpha(t+1) - P_\alpha(t) = \mu \left[ \sum_{\gamma | \alpha \in {\cal A}_\gamma} Q_>(U_\gamma - U_\alpha) P_\gamma(t) - \sum_{\gamma \in {\cal A}_\alpha} Q_>(U_\alpha - U_\gamma) P_\alpha(t) 
\right ].
\ee
As is well known, a sufficient condition for having Eq. (\ref{logit3}) as the equilibrium (long time) solution of this equation is the so-called ``detailed balance'' equation: \footnote{One should
of course keep in mind that the time needed to reach full equilibrium might be very large, or even infinite if there is a phase transition and ergodicity breaking -- 
something that requires the number of states to be infinite.}
\be\label{cond}
\ln \left[\frac{Q_>(U_\gamma - U_\alpha)}{Q_>(U_\alpha - U_\gamma)}\right] = \beta(U_\alpha - U_\gamma), \quad \forall \alpha, \gamma,
\ee
and provided that all choices are reversible, i.e. $\alpha \in {\cal A}_\gamma \Leftrightarrow \gamma \in {\cal A}_\alpha$, $\forall \alpha, \gamma$. 
Clearly,  condition (\ref{cond}) is satisfied when $Q_>(u)$ corresponds to the logistic distribution:
\be
Q_>(u) = \frac{1}{1 + e^{\beta u}},\qquad Q_>(-u) = \frac{1}{1 + e^{-\beta u}} \equiv e^{\beta u} Q_>(u)
\ee
but other distributions may be suitable as well. However, there is no a priori justification for imposing that the distribution of noise differences should have such a property. If the $\epsilon$'s are 
IID (or at least exchangeable), one can deduce in full generality that there is a function $R$ such that:
\be
Q_>(u) = \frac12 - R(u),\qquad R(u) = -R(-u), \qquad R'(u) \geq 0,  \qquad R(u \to \infty) = \frac12. 
\ee
So expanding for small $u$'s one finds that the detailed balance condition is satisfied to first order with $\beta=4R'(0)$, but may be violated for higher orders in utility differences:
\be
\ln \left[\frac{Q_>(-u)}{Q_>(u)}\right] = \ln \left[\frac{1+2R(u)}{1-2R(u)}\right] \approx 4 R'(0) u + \frac23 (R^{(3)}(0) + 8R'(0)^3) u^3 + \dots
\ee
Not much is known about these non-detailed balance models. A natural conjecture is that for this particular class of models, for which the transition rates are functions 
of a utility (or energy) difference between the initial and final state, a quasi-equilibrium state is reached with no large scale currents. The robustness of the phase diagram and of the 
phase transitions to non-detailed balance terms ($u^3$, $u^5$, etc.) is of course a crucial issue. 

The case of binary choice situations with no interactions between agents is clearly much simpler, and can be solved in full generality as:
\be
P(S_i=\pm 1) = \frac12 (1\pm m), \qquad m = 2 R(U_+ - U_-),
\ee
and the introduction of a mean-field interaction leads to a self-consistent equation that generalizes the usual Curie-Weiss equation (Eq. (\ref{CW}) above):
\be
m^* = 2 R(U_+ - U_- + Jm^*).
\ee
It is clear that for generic concave functions $R(u)$ this model still has a second order phase transition when $\beta J=2$, of the type discussed in section \ref{sect2A} (where $2R(u)=\tanh(\beta u/2)$). 
First order
transitions can occur for more exotic shapes of $R(u)$.

But this point of view shows that there is no reason whatsoever to think that the noise parameter $\beta$ should be the same for all agents. Some may be more ``rational'' than others, meaning that 
a distribution of $\beta$'s could be needed -- something that is unusual in physics. The interaction parameter could also be heterogeneous. The criterion for the appearance of hysteresis effects 
is, in the general case, $\langle \beta J \rangle = 2$, where the average is taken over the population. 

\subsection{Non interacting vs. interacting agents, and coordination failure}
\label{sect5B}

It is worth dwelling a little more on the issue of detailed-balance decision rules in the presence of interactions. Call $\alpha_i$ the choice made by agent $i$ and $P(\alpha_1, \alpha_2, \dots,
\alpha_N; t)$ the joint distribution of the choices of the $N$-agents at time $t$, and ${\cal C} = \{\alpha_1, \alpha_2, \dots, \alpha_N\}$ the configuration of choices made by all agents. The 
time evolution of the population is a random walk in the space of choices, and is determined by the transition probabilities $W({\cal C} \to {\cal C}')$ to go from configuration ${\cal C}$ to
configuration ${\cal C}'$ between $t$ and $t+1$. Assume that at each time step, only one agent can change his choice. The set of accessible configurations from ${\cal C}$ is made of all
configurations that differ from ${\cal C}$ by a single agent changing his choice. If every single agent $i$ uses a logit-rule to update his choice from $\alpha_i$ to $\gamma_i$, with possibly $i$
dependent utilities $U^i_\alpha$, one finds for the transition probabilities:
\be
W({\cal C} \to {\cal C}') = \frac{\mu}{1 + e^{\beta (U^i_{\alpha_i}-U^i_{\gamma_i})}}.
\ee
Assuming individual choices are reversible in the above sense, do these transition probabilities $W({\cal C} \to {\cal C}')$ obey detailed-balance? Superficially it seems to be so, but the presence of interactions can
ruin the detailed-balance property. Detailed-balance is obeyed provided there exists a certain function ${\cal H}({\cal C})$ that depends on the choice made by all agents, such that:
\be\label{linking} 
U^i_{\gamma_i} - U^i_{\alpha_i} = {\cal H}(\alpha_1,\alpha_2, \dots, \gamma_i, \dots, \alpha_N) - {\cal H}(\alpha_1,\alpha_2, \dots, \alpha_i, \dots, \alpha_N), \quad \forall i, \quad \forall \alpha,\gamma.
\ee
In this case, one indeed has:
\be
\ln \left[\frac{W({\cal C} \to {\cal C}')}{W({\cal C}' \to {\cal C})}\right] = \beta\left[{\cal H}({\cal C}') - {\cal H}({\cal C})\right],
\ee
such that the long-time equilibrium probabilities are given by the Boltzmann-Gibbs weight:
\be\label{eq-gen}
{\cal P}({\cal C}, t \to \infty) = Z^{-1} e^{\beta H(\left\{ {\cal C} \right\})}.
\ee
When agents are non interacting, in the sense that the utilities $U^i_{\alpha}$ are {\it independent} of the choice made by others, then clearly:
\be
{\cal H}({\cal C}) = \sum_i U^i_{\alpha_i}.
\ee
When there are interactions, this is no longer true, and in fact it is in general {\it not} possible to construct such a function (the consequences of this impossibility will be discussed at the
end of this section).
A simple case where such a construction is possible, for example, is the binary choice situation where $\alpha_i = S_i = \pm 1$, and where the utilities take the following form:
\be
U^i_{S_i} = - \frac{f_i}{2} S_i + \sum_{j \neq i} J_{ij} S_i S_j,
\ee
with $J_{ij} = J_{ji}$ an arbitrary symmetric matrix. In this case, one can check by inspection that ${\cal H}$ is given by:
\be
{\cal H}(\{ S_\ell \}) = - \sum_i \frac{f_i}{2} S_i  + \frac12 \sum_{i \neq j} J_{ij} S_i S_j.
\ee
One sees on this example that $\cal H$ is not the sum of individual utilities. In fact:
\be
{\cal H}(\{ S_\ell \}) -  \sum_i U^i_{\alpha_i} = - \frac12 \sum_{i \neq j} J_{ij} S_i S_j.
\ee				      
The difference between these two quantities can have important consequences, as pointed out in the beautiful paper of Grauwin et al. \cite{Grauwin}. 
They consider a variant of the Schelling model of segregation, where each 
agent $i$ is sensitive to the density $\varrho \in [0,1]$ of the neighbourhood in which he/she is living. Agents do not like living in derelict 
neighbourhoods ($\varrho \simeq 0$) nor in overcrowded neighbourhoods ($\varrho \simeq 1$). Their incentive to live in a neighbourhood situated around 
$\vec r$ is a certain function $\upsilon[\varrho(\vec r)]$ that has a maximum at $\varrho=1/2$; for example $\upsilon(u \leq 1/2) = 2u$ and  
$\upsilon(u > 1/2) = 2 - 2u$. The global utility is a functional of the density field $\varrho(\vec r)$ and is given by:
\be
U_G \left(\{\varrho(\vec r)\}\right) = \int_{L^2} d\vec r \varrho(\vec r) \upsilon[\varrho(\vec r)],
\ee
where $L$ is the linear size of the city. Suppose that the size $N$ of the population is such that $N/L^2 = 1/2$. It is in principle possible in this 
case to make everybody happy and reach the absolute maximum of the global utility of the population by having all neighbourhoods at $\varrho =1/2$, 
in which case $U_G^* = N \upsilon_{\max}$, with $\upsilon_{\max}=\upsilon(u = 1/2) = 1$. 

Now, what will really happen if agents choose their neighbourhood based on their own utility? More precisely, let us assume that each agent can 
move from $\vec r$ to $\vec r^{\, \prime}$ with a probability given by the logit-rule, i.e.:
\be
W(\vec r \to \vec r^{\, \prime}) = \frac{\mu}{1 + e^{\beta(\upsilon[\varrho(\vec r)] - \upsilon[\varrho(\vec r^{\, \prime})])}}.
\ee
The function ${\cal H}$ alluded to above is also a functional of $\varrho(\vec r)$, and according to the general principle expressed by Eq. (\ref{linking}),
it should be such that, for all fields $\varrho(\vec r)$:
\be
\frac{\delta {\cal H}}{\delta \varrho(\vec r^{\, \prime})} - \frac{\delta {\cal H}}{\delta \varrho(\vec r)} =  \upsilon[\varrho(\vec r^{\, \prime})] 
- \upsilon[\varrho(\vec r)];
\ee
where we have assumed that the elementary area of a neighbourhood is normalized to one. The solution is:
\be
{\cal H}[\{\varrho(\vec r)\}] = \int_{L^2} d\vec r \left[\int_0^{\varrho[\vec r]} {\rm d}\varrho' \,  \upsilon[\varrho'] + A \varrho[\vec r]\right],
\ee
where $A$ is an arbitrary constant. Therefore, unless
$\upsilon$ is a constant, independent of $\varrho$ (i.e. for non interacting agents), ${\cal H} \neq U_G$.

Now, the equilibrium probability is given by the Boltzmann-Gibbs measure: \footnote{We neglect here an entropic term, which is small when $\beta \to \infty$, see  \cite{Grauwin} for
details.}
\be
P[\{\varrho(\vec r)\}] \propto \exp\left[\beta \int_{L^2} d\vec r \left(V(\varrho[\vec r])-\lambda \varrho[\vec r]\right)\right],
\ee
with $V'(.)=\upsilon(.)$ and $\lambda$ is a Lagrange parameter ensuring that the average density is fixed to a certain $\overline \varrho$.  
The most likely density configuration, for $\beta \to \infty$, is such that $V'=\lambda$, $\forall \vec r$. For the above symmetric 
piecewise linear choice for $\upsilon(u)$, one finds that the local density must be equal to $\varrho=1/2 - \delta$ with probability $p$, and $\varrho=1/2 + \delta$ with probability $1-p$, 
with $p$ such that $\overline \varrho = 1/2 + \delta(2p-1)$, and $\delta \in [-1/2,1/2]$ such that:
\be
I(\delta)= p \left(\frac12 - \delta\right)^2 + (1-p) \left(\frac12 + 2 \delta - (\frac12 + \delta)^2\right) =
\frac14 + (\overline \varrho - \frac12)(\delta - 1) 
\ee
is maximized. In the case where $\overline \varrho > \frac12$, the maximum is realized for $\delta = 1/2$, and when $\overline \varrho < \frac12$, it
is realized for $\delta = -1/2$, but both solutions correspond to complete segregation, i.e. completely full ($\varrho=1$) or completely empty 
($\varrho=0$) neighbourhoods! (The case where $\overline \varrho = 1/2$ is degenerate, but any infinitesimally small departure from this case leads 
to segregation). Therefore, letting people optimize their own utility, to the detriment of the global well-being, leads to a disastrous situation. This is
a stunning illustration of a case where Adam Smith's invisible hand totally fails to bring the society to a global optimum.

Now, as proposed in  \cite{Grauwin}, coordination can be improved by making individuals conscious of the common good when they make their choice, by
adding to the change of individual utility a term proportional to the extra cost in the global utility, i.e. change the logit-rule according to:
\be
\Delta \upsilon =\upsilon[\varrho(\vec r^{\, \prime})] - \upsilon[\varrho(\vec r)] \longrightarrow \Delta \upsilon
+ \zeta \left[\Delta U_G - \Delta \upsilon\right].
\ee
Since $U_G$ is already a functional of the density field, one can immediately see that the function ${\cal H}_\zeta[\{\varrho(\vec r)\}]$ adapted
to the new choice update rule is given by:
\be
{\cal H}_\zeta = (1 - \zeta) {\cal H} + \zeta U_G.
\ee
Correspondingly, the equilibrium probability is now replaced by:
\be
P_\zeta[\{\varrho(\vec r)\}] \propto \exp\left[\beta \int_{L^2} d\vec r \left\{(1-\zeta)V(\varrho[\vec r])+\zeta\varrho[\vec r]\upsilon(\varrho[\vec r])
-\lambda \varrho[\vec r]\right\}\right].
\ee
The most likely configuration can again be computed. One finds that it becomes the socially optimal solution $\varrho[\vec r] \equiv 1/2$ and $U_G^*/N =\upsilon_{\max}$
as soon as $\zeta > \zeta_c = 1/3$. As stressed by Grauwin et al., this model shows that {\it the optimization strategy based on purely individual dynamics
fails, illustrating the unexpected links between micromotives and macrobehavior}. Making individuals aware of the costs of their choice for the
society is, in some cases, useful to nudge the system into cooperation/coordination \cite{Nudge}. 

Note that the results of Grauwin et al. hold provided the decision rules obey detailed-balance. However, detailed-balance is, in the general case of interacting agents, 
the exception rather than the rule. Very few 
general results are known when detailed-balance is violated, see e.g. \cite{Jona-Lasinio,CK,Biroli}, see also \cite{Burioni} in the present context. 
Of course, some equilibrium will be reached after a long time, but there is no explicit 
construction of the equilibrium probability, similar to Eq. (\ref{eq-gen}) above. One knows that in the general case, equilibrium {\it currents} 
are present. This is due to the fact that the probability to cycle `clockwise' around a closed loop in decision space is different from the probability to 
cycle `counter-clockwise' -- except when detailed-balance is satisfied. Studying and interpreting these currents in a socio-economic context would be very interesting, in 
particular because these currents are known to considerably speed up the equilibration time in certain cases \cite{SG_asym,CK}, preventing the system from getting ``trapped'' in 
deep maxima of the function ${\cal H}$ (the equivalent of deep energy wells in physics). For example, a natural extension of the model of Grauwin et al. that would 
exhibit equilibrium currents and (perhaps ?) mitigate segregation, would be quite appealing.

\section{Imitation of the past and more instabilities}

\subsection{Memory \& Habit formation}
\label{sect6A}

Any realistic model of decision should in fact take into account that the utility/incentive of a given choice cannot be thought of as time independent. For example, it is
hard to know how useful or interesting choice $\alpha$ is without having tried it at least once (think about restaurants, for example). A simple way to model this would be to
to assume that the perceived utility of choice $\alpha$, $U_\alpha^i$, is initially blurred by some estimation error $\epsilon^i_\alpha$ that decays to zero as the total time
spent choosing $\alpha$ grows; for example \cite{YS}:
\be
\epsilon^i_\alpha(t) = \epsilon^i_\alpha(t=0) e^{-\Gamma \varphi_\alpha t},
\ee
where $\varphi_\alpha$ is the fraction of the time during which $\alpha$ was chosen. In effect, this should be similar to having an effective parameter $\beta(t)$ that increases over time. 
This is the equivalent, in physics, of an annealing process where the temperature $T=1/\beta$ slowly decreases with time. 

As the relative utility of choices becomes better known to agents as time evolves, a second, more important effect might kick in: habit formation or ``stickiness''. Often, past choices become more valuable
only because they happened to be chosen. This can be for good reasons, like creating a loyalty relationship (such at the Marseilles fish market \cite{Kirman_fish,WKH}), or because going to the
same doctor makes that doctor know your personal problems better, etc. But it can also be because of high risk aversion (maybe another choice is better, but maybe it is much worse, so better stick
to what I have) or sheer intellectual laziness. Hotel and restaurant chains are based on this strong universal principle: people often tend to prefer things they know. 

This can be formalized as follows: define an indicator function $\theta_{\alpha}(t')$, which is equal to $1$ if choice $\alpha$ is made at time $t'$, and zero otherwise. The utility of choice 
$\alpha$ at time $t$ for a given agent is written as:
\be
U_\alpha(t) = U_\alpha(0) + \sum_{t' < t} K_\alpha(t-t') \theta_\alpha(t'),
\ee
where $U_\alpha(0)$ can be thought of as the `true' utility of choice $\alpha$ and $K_\alpha(\tau)$ is a certain memory kernel, which measures how much of the far away past is affecting the utility today. 
When $K_\alpha > 0$, choices are self-reinforcing.\footnote{This is similar to reinforcement learning rules, see e.g. \cite{MG,Farmer_new} for some references in the present context.}  This model,
with an exponentially decaying memory $K_\alpha(\tau) = \varkappa q^\tau$, was introduced in \cite{Nadal_Kirman} to model loyalty formation between buyers and sellers in the Marseilles fish market. 

Suppose a somewhat degenerate case where initially $U_\alpha(0) \equiv 0$, i.e. all choices are {\it a priori} equally good. What happens at long times? Is loyalty a long-term emergent property due to `imitation of the past', 
similar to the appearance of `winners' in the artificial cultural market of Salganik et al. due to imitation of past choices of others (\cite{Salganik} and section \ref{sect4D} above)?
The analysis proposed in \cite{Nadal_Kirman} is based on a mean-field argument. If the utilities $U_\alpha(t)$ do not change too fast, equilibrium can be assumed to hold at each time step, 
leading to:
\be
P_\alpha(t) \approx  Z(t)^{-1} e^{\beta U_\alpha(t)},
\ee
where $Z(t)$ is the time dependent normalisation. Now, replacing the history dependent $U_\alpha(t)$ by its average leads to a self-consistent equation for $P_\alpha(t)$. Assuming that $P_\alpha(t)$ converges at large times towards $P_\alpha^*$,
these should be given by the solution of:
\be
\ln P_\alpha^* = - \ln Z^* + \frac{\beta \varkappa q}{1-q} P_\alpha^*
\ee
Clearly, $P_\alpha^*=1/M$, $\forall \alpha = 1, \dots, M$ is a always solution, corresponding to the absence of loyalty formation -- if $\alpha$ denotes the seller in a fish market, it is a situation
where buyers tend to change their seller every week, and no long term relationship arises. However, this fully diversified solution becomes unstable when $\beta \varkappa q > M(1-q)$, and the solution becomes 
`polarized', i.e. a habit, loyalty or addiction spontaneously appears beyond some reinforcing strength $\varkappa$, assuming the memory time (set by $q$) is fixed. Note that this is an example of 
a {\it condensation} phase transition which occurs even with a single agent, since we have not introduced any interaction between agents in the model, only a memory feedback. When the `true' utilities 
are different, the same effect can exist: habit can lead to a persistent choice of suboptimal decisions.

This is a very nice effect, but unfortunately the mean-field argument is flawed in this particular case. In order to see why and to possibly rescue the result, let us focus to the binary decision
case, where $\alpha(t) = S(t) = \pm 1$. In this case, only the difference $U_+(t) - U_-(t)$ matters, which can be written as:
\be
U_+(t) - U_-(t) = U_+(0) - U_-(0) + \sum_{t' < t} K(t-t') S(t').
\ee
Assuming $U_+(0)=U_-(0)$ and using the logit decision rule Eq. (\ref{logit1}) with $\mu=1$, one establishes that the probability to observe a certain history of choices $\{ S_t \}_{0 \leq t \leq L}$ is given by:
\be
P\left[\{ S_t \}_{0 \leq t \leq L}\right] = \frac{\exp\left[{\beta \sum_{0 \leq t \leq L} S(t) \sum_{t' < t} K(t-t') S(t')}\right]}
{\prod_{0 \leq t \leq L} 2 \cosh \left[{\beta \sum_{t' < t} K(t-t') S(t')}\right]}.
\ee
But this is similar to the Boltzmann-Gibbs weight for a spin configuration of the one dimensional Ising model, with an interaction matrix $K(t-t')$ that depends on the ``distance'' $t-t'$ between the spins. The difference comes from the denominator, absent in the Ising model; but one can argue that if there is no phase transition in the Ising model, there is no transition in the above model either. \footnote{More rigorous work is needed on this whole issue; in particular on 
whether the existence of a transition in the Ising model is sufficient to ensure loyalty formation.}
Quite a lot is known about the one dimensional Ising model in the limit $L \to \infty$ \cite{Thouless,Dyson}. In particular, it can be proven that when $K(\tau)$ decays faster than $\tau^{-2}$ when $\tau \to \infty$, the system is always in the 
high temperature, disordered phase, where the magnetisation is zero. This means that the two choices remain equally probable at long times; there is no loyalty formation and ergodicity is not broken. Still, it is 
known that the correlation length of the Ising model (and thus the correlation time in the present version) grows exponentially fast with the interaction strength. For example, when the memory decays 
exponentially as in the model of \cite{Nadal_Kirman}, one finds that the typical time during which an agent remains faithful is given by $\exp[2\beta \varkappa q/({1-q})]$ when $\beta \varkappa \gg 1$. 
The transition found in \cite{Nadal_Kirman} is thus in fact only a crossover between a fickle phase and a sticky phase. However, for longer ranged memory kernels, a true phase transition takes 
place. Suppose $K(\tau)$ decays asymptotically as $\tau^{-\nu}$ with $1 < \nu \leq 2$, then there is indeed a critical value of $\beta$ beyond which loyalty sets in the sense that the probability to choose $+1$ or $-1$ is not 
$1/2$, $1/2$ anymore. The  marginal case $\nu=2$ is in fact special and very famous: it corresponds to the Thouless-Anderson-Yuval model with a long history in condensed matter physics \cite{Thouless,Anderson_Yuval}.
When $\nu < 1$, loyalty appears for any non-zero value of $\beta$. Similar conclusions are expected to hold for multiple choice situations as well. The case where memory adds to interaction between agents was 
recently considered in the context of the Ising model in \cite{Ising_memory}.

It would be interesting to try to determine empirically whether people's memory of past choices decays exponentially or as a power-law. Power-law memory is well documented in many human activities, including financial 
markets, see e.g. \cite{Barabasi,Deschatres,Chicheportiche} and references therein. We see from the above model that this question might be quite relevant for the appearance of persistent choices and decisions, or 
{\it self-fulfilling prophecies}, which are very closely related to the above scenario, and to which we devote the next section.

\subsection{Conventions and self-fulfilling prophecies}
\label{sect6B}

As we mentioned above, habit formation is due to `imitation of the past' rather than `imitation of the peers'. In fact, both effects can be present and reinforce each other. \footnote{Interesting 
situations could occur when these two effects are in conflict, for example when overcrowding or saturation effects prevent full condensation.}
We want to pursue this further and consider other theoretical scenarios of feedback of the past on current decisions, that may lead to self-fulfilling prophecies \cite{Woodford} and endogenous crashes \cite{Risk}.
This of course opens the whole field of {\it learning models}, much emphasized in e.g. \cite{Arthur,Brock_Hommes,MG,Farmer_new}, that is much too big to be fairly covered in this paper. 

In the context of repeated decisions (such as investing in financial markets), a common temptation is to compare
the present situation with similar situations from the past, and posit that what already happened is
more likely to happen again. As Brian Arthur puts it: {\it As the situation is replayed regularly, we
look for patterns, and we use these to construct temporary expectational models or hypotheses
to work with} \cite{Arthur}. Often some plausible story is given to understand why such a pattern should exist.
This convinces more participants that the effect is real, and their resulting behaviour reinforces
(or even creates) the effect: this is a self-fulfilling prophecy. A large consensus
among economic agents about the correlations between a piece of information and the market
reaction can be enough to establish these correlations. Such a `condensation' of opinions leads to
what Keynes called a convention \cite{Keynes,Orlean_Book}, a commonly shared representation of the world on which uncertain agents can rely on to
make a decision. However, as he further noted, {\it a conventional valuation which is established as the outcome of the mass psychology of a large number of ignorant individuals
is liable to change violently as the result of a sudden fluctuation of opinion due to factors which do not really make much difference...}.

As a striking example of a sudden change of convention, we show in Fig. 5 the correlation between bond markets and stock markets as a function of time. This correlation was
positive before 1997, because the common lore suggests that low long term interest rates should favor stocks. Since bonds move opposite to rates, an increase in bond price should 
trigger an increase in stock prices. But 
another story  suddenly took over in the midst of the 1997 Asian crisis: a fall in stock markets triggered
an increased anxiety of the operators who sold their risky paper and bought non-risky government
bonds. This so-called Flight to Quality then became the dominant pattern. There are many other anecdotal examples of such conventions in financial markets: indicators or news items (inflation, 
payrolls, etc.) that everybody pays attention to at one moment in time, before collectively deciding it is irrelevant after all.

 \begin{figure}
      \begin{center}
      \epsfig{file=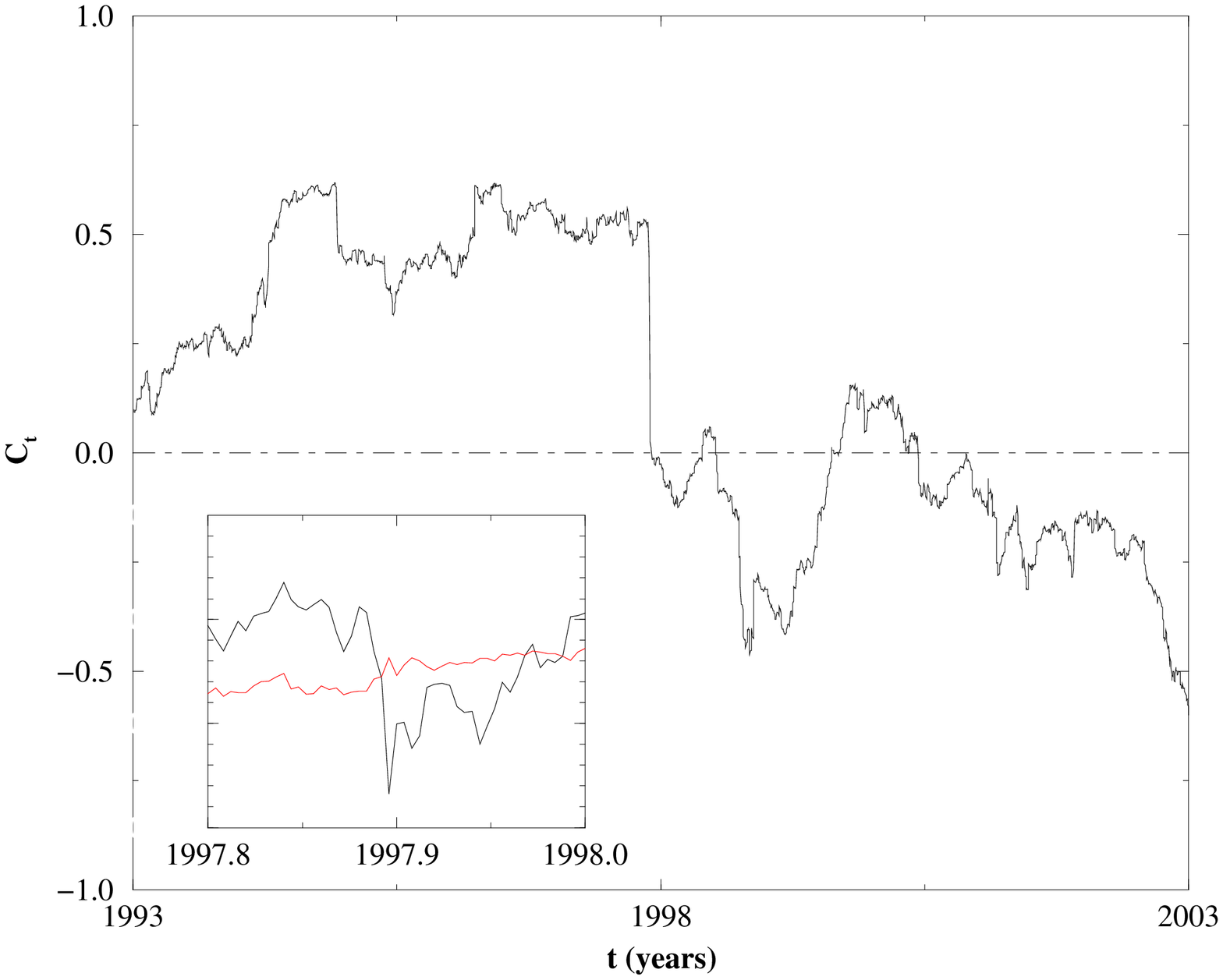,width=5cm,angle=270}\hspace{1cm}
      \epsfig{file=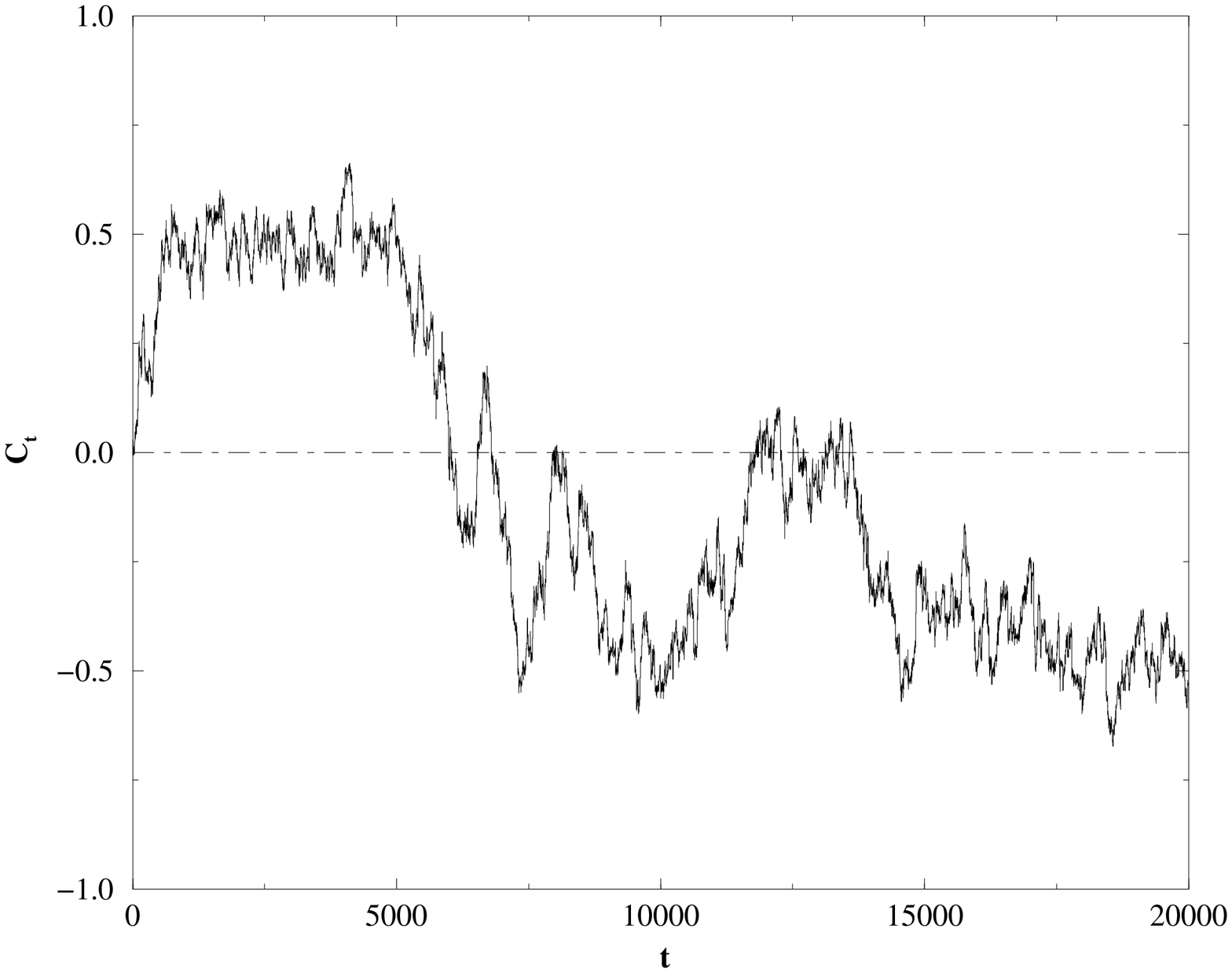,width=5cm,angle=270}
      \end{center}
      \caption{Left: Normalized correlation between the Dow-Jones daily
      returns and the daily 
      returns of 
      a U.S. bond index with 7 to 10 years bonds, computed with an exponential 
      moving average of $100$ days. Note the 
      convention change occurring at the end of 1997.
      Inset: Zoom on the evolution of the Dow-Jones and the bond index in the last
      quarter of 1997, during the Asian crisis. 
      Right: One characteristic trajectory of the dependent correlation $c_t$ in the model defined by Eqs. (\ref{Ct},\ref{price}), for $g=1.2$
      and $1-q=0.01$.}
      \label{Fig5}
 \end{figure}

The sudden shift seen in Fig. 5 is quite remarkable. A model that could explain such a change in convention was proposed in \cite{WB}. 
Call $p_t$ the (log-)price of a certain asset at time $t$, and $\delta p_t$ the  
return between $t$ and $t+1$. Some agents base their strategy for buying or selling on the observation
of the temporal change of a certain index ${\cal I}_t$, which might be a financial index or 
an economic indicator (for example, dividends, interest rates, inflation, confidence, unemployment, etc.), or even the price $p_t$ itself
or the price of another asset. Denote by $\delta {\cal I}_t$ the change of this indicator between $t$ and $t+1$, normalized to have zero mean and unit variance.
Suppose that it is reasonable that there should exist a causal correlation between $\delta {\cal I}_t$ and
$\delta p_{t+1}$, i.e. that the correlation coefficient between the two time series is:
\be
   E\left[\delta {\cal I}_t \delta p_{t+1}\right] \equiv c. 
\ee
It is well known that the best prediction (in a quadratic 
sense) of $\delta p_{t+1}$ knowing $\delta {\cal I}_t$ is given by 
   \be 
   \delta p_{t+1}^* = {c} \delta{\cal I}_t. 
   \ee
Now, imagine that there are two types of agents, those who act based some information uncorrelated with ${\cal I}_t$ and those 
who try to take advantage of the possible correlations between $\delta {\cal I}_t$ and $\delta p_{t+1}$. 
Since the true, fundamental value of the correlation $c$ is in fact not known (and might actually be zero!), agents 
of the second type attempt to learn this value from past history. Similar to the above model of \cite{Nadal_Kirman}, let 
us assume that the memory of the past is decaying exponentially, leading to an {\it estimated} value of $c$ 
at time $t$ given by:
   \be\label{Ct}
   c_t = \frac{1-q}{q} \sum_{t'=-\infty}^{t-1} q^{t-t'} (\delta I_{t'} \delta p_{t'+1}),
   \ee
where $q$ sets the memory time of the exponential moving average. Equation (\ref{Ct}) is equivalent to the following Markovian update of the estimated correlation
$c_t = q c_{t-1} + (1-q) \delta I_{t-1} \delta p_{t}$.

Agents using the information based on ${\cal I}_t$ then decide 
to buy (or sell) a quantity $\Phi_t$ that is a (odd) function of the predicted next return $\delta p_{t+1}^* \propto c_t \delta {\cal I}_t$.
One expects on general grounds that the demand function $\Phi_t$ is linear for small arguments and saturates for large arguments. The small $c$ expansion 
should thus read: 
\be
\Phi_t = a c_t \delta {\cal I}_t - b (c_t \delta {\cal I}_t)^3 + O(c^5),
\ee
where $a,b$ are positive coefficients. These orders add to the demand $\Omega_t$ of all other agents, who use a different type of information. Assuming  
the aggregate order imbalance has a linear impact \footnote{Whereas the impact of individual orders in strongly concave, the impact of the aggregate order imbalance is, to a 
good  approximation, linear. See \cite{marketimpact,Toth} for a long discussion of this issue.}, we finally end up with the following expression for the {\it realized} 
price change at $t+1$:
\be\label{price}
\delta p_{t+1} = \lambda \left[ \Omega_t +  a c_t \delta {\cal I}_t - b c_t^3 \delta {\cal I}_t^3 \right] + O(c^5),
\ee
where $\lambda$ is the impact coefficient, relating volume imbalance to price changes: $\delta p = \lambda \Phi$. We will define the feedback strength parameter as $g \equiv a \lambda$, which plays 
a central role in the model, and assume that the `true' long time average correlation $c$ is zero, i.e. there is in fact {\it no} information in ${\cal I}_t$. 

Combining Eqs. (\ref{Ct},\ref{price}) and taking a continuous time limit (justified in the limit of 
long memory kernels, $q \to 1$), one finally finds the following Langevin equation for $c$ (see \cite{WB} for details):
\be
    {dc} = - \frac{dV}{dc} {dt} +  d\xi,
\ee
where $d\xi$ is a Brownian noise (related to $\Omega$) of variance $1-q$ and $V$ is an effective potential given by:
\be
    V(c) = \frac{1}{2} (1-g) c^2 + \frac{1}{4} b' c^4, 
\ee
with $b'$ another positive coefficient \cite{WB}.

The equilibrium distribution of estimated correlations $c$ is then easily found to be given by:
\be\label{Bolz}
    P(c) = \frac{1}{Z} \exp \left(-\frac{2 V(c)}{1-q}\right),
\ee
 For $g < 1$ the potential has an absolute minimum at $c=0$, whereas for 
$g > 1$, $c=0$ becomes a local maximum and two stable minima appear for 
$c=\pm c^*=\pm \sqrt{(g-1)/b'}$. 
Therefore, for $g < 1$ (weak feedback), $P(c)$ is unimodal and has a maximum at $c=0$, reflecting the 
underlying reality. When feedback is stronger ($g > 1$), however, the most probable values for $c$
are $\pm c^*$. This means that for strong feedback, a non-zero fictitious 
correlation between the price and the indicator spontaneously appears -- this is the so-called sunspot effect \cite{Woodford}. 
The correlation can be either positive or negative, corresponding to the two 
possible `conventions'. However, the long time average of correlation is still zero even when $g > 1$, since $c$ randomly flips
between $\pm c^*$. But the time needed to do so is however exponentially large in the memory time of the agents, $1/(1-q)$. 
Hence non-zero estimated correlations can therefore persist for very long times, as in the exponential memory model of \cite{Nadal_Kirman}.
One therefore expects to observe long periods during which $c \approx +c^*$, until the convention reverts quite abruptly towards $c \approx -c^*$, 
as observed for the bond-index correlation shown in Fig. 5, where we also show a numerical simulation of the model corresponding to $g=1.2$ and $q=0.99$.

Finally, suppose now that there indeed exists a {\it small} objective correlation $c_0 \ll 1$ between 
$\delta p_{t+1}$ and $\delta I_t$, justified by some real economic mechanism relating the two quantities. One can show \cite{WB} that this 
is accounted for by adding  a linear contribution to the effective potential $V(c)$:
\be
V(c) \longrightarrow V(c) - c_0 c.
\ee
For $g < 1$, the most probable value of $c$ is $c_0/(1-g)$ ($=c_0$ for $g=0$, as it should in the absence of any feedback). Therefore, $c$ is of the same
order as its true cause whenever $g < 1$. However, in the limit $g \to 1^-$, the 
apparent correlation becomes much larger than its true cause: the sensitivity of the market to external information is anomalously amplified by the impact of traders 
using that information. For $g > 1$, 
the amplitude of the apparent correlation is totally unrelated to that of the true correlation, although the {\it sign} of the correlation reflects the underlying economic reality. One
describes here a typical example of overreaction to news leading to excess correlations that are well documented in the literature (see e.g. \cite{DeBondt}). For example, the 
correlations between the stocks belonging to an index and the index itself are too strong to be explained by the intrinsic correlations between the stocks \cite{Shiller}. These increased
correlations seem to reflect a self-induced ``risk-on/risk-off'' mode (see \cite{Marsili_Raf2} for similar ideas).

\subsection{Self-induced bubbles and collapse}
\label{sect6C}

In the previous model, agents were trying to learn correlations to make a forecast of the next price change. But among the most pervasive biases in human perception is the detection of trends (or the illusion of trends). 
The human eye seems to be programmed to make a ``fast and frugal'' linear regression, and extrapolate the line into the future (see \cite{Hommes} for experimental evidence of this forecasting rule). 
Humans (and of course, animals) are also very good at detecting danger. These two basic features can be minimally encapsulated in the following ``mood indicator'' ${\cal M}$ \cite{BC,Joao}:
\be
{\cal M}_{t} = a {\cal T}_t - b {\cal R}_t,
\ee
where ${\cal T}_t$ is a trend indicator and ${\cal R}_t$ a risk indicator, and $a,b$ again two positive parameters. The $a$ term describes the unrelenting propensity of humans to follow trends, while the $b$ term can 
be thought of as a risk aversion term, prompting investors to sell when the perceived risk becomes too large.
For simplicity, one assumes that both  ${\cal T}_t$ and ${\cal R}_t$ are constructed as above as exponential moving averages of respectively the past 
price changes and the past squared price changes:
\be
{\cal T}_t = \frac{1-q}{q} \sum_{t'=-\infty}^{t-1} q^{t-t'} \delta p_{t'}, \qquad  {\cal R}_t = \frac{1-q}{q} \sum_{t'=-\infty}^{t-1} q^{t-t'} [\delta p_{t'}]^2.
\ee

Now, as argued in \cite{BC} and more recently in \cite{Toth}, supply and demand in financial markets cannot be instantaneously matched. This is due to the fact that the immediate outstanding liquidity in the
order book, or provided by market makers, is a tiny fraction (say 1 \% or less) of the total daily volume. So in fact there is always a certain amount of pending demand or pending supply, which was called 
``latent liquidity'' in \cite{Toth}. The idea of \cite{BC} was to write the following dynamical equation for the latent order imbalance $\Phi_t$:
\be
\Phi_{t+1} - \Phi_t = - \mu \Phi_t + {\cal M}_{t} + \Omega_t
\ee
where the first term comes from the fact that a well-functioning market allows to the order imbalance to relax to zero on a time scale $\mu^{-1}$ (through execution and cancellations), whereas the second term describes
the arrival of new buy orders or sell orders, triggered by the mood indicator ${\cal M}_{t}$. The extra random noise $\Omega_t$ is coming from exogenous events or from the behaviour of agents following different indicators. 
The price evolution equation is still given by the simplest linear model: on average, the latent order imbalance $\Phi_t$ biases the price upwards or downwards as $\delta p_{t} = \lambda \Phi_t$. 

There are two time scales in the model: $\mu^{-1}$ (related to the liquidity of the market) and $(1-q)^{-1}$, the memory time over which trends and risk are computed. The analysis of the model can be made in 
both limits $\mu \ll (1-q)$ and $\mu \gg (1-q)$, with similar conclusions, see \cite{BC}. In order to keep the discussion as simple as possible, we focus here on the case of extreme short memory, i.e. $q \to 0$, 
in which the dynamics of the price is given by:
\be
p_{t+2} - 2 p_{t+1} + p_t = \lambda (\Phi_{t+1} - \Phi_t) = - \mu (p_{t+1} - p_t) + a \lambda (p_{t} - p_{t-1}) - b\lambda (p_{t} - p_{t-1})^2 + \lambda \Omega_t.
\ee
This evolution is easier to interpret in the continuum time limit. Set $v = dp/dt$ (the price ``velocity''); then up to small terms the above equation reads:
\be
\frac{dv}{dt} = - \frac{dW}{dv} + \lambda \Omega_t, \qquad W(v) := (\mu - \lambda a) \frac{v^2}{2} + \lambda b  \frac{v^3}{3}
\ee
which describes the overdamped evolution of the position $v$ of a fictitious particle in a cubic potential $W(v)$,  in the presence of noise. 

When trend following effects are small, or the market liquid enough ($\lambda a \ll \mu$), the potential $W(v)$ has a minimum at $v=0$, which translates into a random-walk regime for the price itself (at least 
on time scales $\gg \mu^{-1}$, see \cite{BC}). But $W(v)$ also has an unstable maximum for a negative value $v^*$, beyond which the potential plummets
to $- \infty$. This describes a risk-aversion induced crash: when several small negative shocks $\Omega_t$ accumulate by accident, the risk-aversion term $-b {\cal R}_t$ induces further selling pressure. This unstable
feedback loop leads to a crash once the tipping point $v^*$ is reached. But for the crash to occur, an ``energy barrier'' $W(v^*)$ has to be crossed, a situation that we have encountered  in 
the above sections (see sections \ref{sect2B} and \ref{sect6B}). The time needed to cross this barrier is, again, exponentially large (in $1/b^2$): for small enough risk-aversion, these crashes are very rare events. 

Now, the interesting regime is when trend following effects become strong, i.e. when $\lambda a > \mu$ -- meaning that the market cannot fully digest the pressure imposed by trend followers. 
In this case, the potential $W(v)$ has a stable minimum for $v^* = (\lambda a - \mu)/b > 0$ around which 
$v(t)$ oscillates. This means that a trend has appeared: the (log-)price is a random walk with drift $v^*$ and thus  on average increases linearly with time, as $p_t - p_0 \approx v^* t$. 
This is a self-induced speculative bubble, which persists because of the trend following tendency of investors, and the inability of the market to absorb the demand quickly enough. 
 
Within the context of the model, there are two possibilities for the bubble to burst. One is a risk-aversion crash, as discussed above, where an accidental fall of the price, driven by the noise term $\Omega_t$, brings 
the system close to the ``cliff'', i.e. the maximum of $W(v)$ now situated at $v=0$, beyond which the crash unfolds. The second possibility is the presence of investors who believe in a fundamental price level, $p^*$. It 
is reasonable to model this class of investors as contributing to $\Omega_t$ in proportion to the difference between the actual price $p_t$ and the fundamental price $p^*$. Clearly, a large discrepancy is an incentive
to bet on a probable reversion of the price towards $p^*$. As the bubble grows, the price $p_t$ moves further and further away from the fundamental value: $p_t - p^* \approx v^* t$. 
In a first approximation, this effect adds to $W(v)$ a time-dependent linear contribution:
\be
W_{eff}(v) \approx  \lambda \phi \, v^* t \, v + (\mu - \lambda a) \frac{v^2}{2} + \lambda b  \frac{v^3}{3},
\ee
where $\phi > 0$ measures the strength of fundamental investing. Interestingly, the minimum of this potential (corresponding to the self-sustained bubble) only exists as long as:
\be\label{tc}
v^* t < \frac{(a - \mu/\lambda)^2}{4b\phi} \longrightarrow t < t_c = \frac{a \lambda - \mu}{4 \lambda^2 \phi}.
\ee
When the price is too high, the self-sustained bubble becomes unstable and a crash  occurs, with $v$ moving to large negative values. Note that, as expected, $t_c$ becomes small if 
fundamental investing becomes dominant (i.e. $\phi$ large).

Note also that Eq. (\ref{tc}) gives only a rough order of magnitude of the critical time $t_c$ beyond which the minimum of the potential
ceases to exist. The real mathematical problem requires one to deal with the following non-linear stochastic differential equation:
\be
\frac{d^2p}{dt^2} = (\lambda a - \mu) \frac{dp}{dt} - b \lambda \left(\frac{dp}{dt}\right)^2 - \lambda \phi (p - p^*) + \lambda \Omega_t.
\ee
However, the simplified treatment in terms of an effective, time dependent potential $W_{eff}(v)$ suggests that the crash is associated to the same generic saddle-node bifurcation as the one recently studied in \cite{Pomeau} in the 
context of earthquakes and material creep. \footnote{In fact, the {\sc rfim}  has also been related to the physics of earthquakes and material creep, see \cite{Sethna_nature,Dahmen}. It would be interesting to make a detailed connection with the work of ref. 
\cite{Pomeau}.} The authors of \cite{Pomeau} suggest that some precursor effects, like a shift of the power-spectrum of the fluctuations to lower frequencies, could be used as a signal of the incipient instability. Whether these ideas
can be transposed to financial instabilities is of course highly conjectural. 

The above model, although highly stylized, seems to capture several interesting facets of the dynamics of financial markets: trends feed trends and risk-aversion can lead to crashes. All these effects come about because 
past events feed back onto present decisions. The model also illustrates the competition between trend following strategies and fundamental strategies, that leads to bubbles and crashes, and intermittent dynamics --
as is also the case for several agent-based models of financial markets with the same ingredients (see e. g. \cite{Lux_Marchesi,GB}, and for recent reviews, \cite{Hommes_review,Stauffer_review,Pietronero,Abergel}). 

\section{Conclusion} 

Financial and economic history is strewn with bubbles and crashes, booms and busts, crises and upheavals of all sorts \cite{Ferguson}. Time series of prices exhibit discontinuities on all scales \cite{Gabaix,Risk}. Often, these major events seem unrelated 
to any fundamental news (see \cite{Summers,Fair,Joulin}), or at least look incommensurate with any reasonable proximate cause. Equilibrium models, often expressed in terms of the behaviour of a single rational `representative
agent', are perhaps adequate to describe small excursions around a stable situation, but are unfit to explain, or even accommodate, major disruptions. The big problem is that these crises are anything but rare \cite{Thistime}: 
{\it the economy is not a ship sailing on a well-defined trajectory which occasionally gets knocked off course} \cite{Kirman_Book}. 
As stressed by Kirman, {\it the economic crisis is a crisis for economic theory} \cite{Kirman_crisis}. However, having declared the old theory dead and buried is not enough. One is faced with a crucial problem: providing a 
viable alternative. Until such an alternative theory is available and to some extent operational, all professional economists (professors, researchers, experts of all kind, policy makers) will prefer to stick with the previous paradigm. After all,
Ptolemy's epicycles remained for many years better than Newton's theory at predicting the motion of planets.

Still, the idea that {\it heterogeneities and interactions} are crucial to understand crises should be taken seriously both by researchers and by policy makers. The aim of this paper was to review recent efforts to include heterogeneities and interactions in models of decision. We have argued that the {\sc rfim} provides a unifying framework to account for many collective socio-economic 
phenomena that lead to endogenous ruptures and crises. Interestingly, the {\sc rfim} has been proposed as a prototype model to describe many physical crises: magnetic avalanches, earthquakes, solar flares, material failures, etc. 
\cite{Sethna_nature}. In all these examples of ``complex systems'', small causes may lead to large effects. The analogy between these physical phenomena and economic/financial turbulence is enticing, at least at a superficial level. In this respect, it is certainly rewarding to set up models that can capture potentially destabilising 
self-referential feedback loops (induced either by {\it herding}, i.e.  reference to peers, as in sections \ref{sect2A}, \ref{sect3B} or {\it trending}, i.e. reference to the past, as in sections \ref{sect6A}, \ref{sect6B} and \ref{sect6C}), and
that can reproduce some of the phenomenology missing in the standard models. However, the real challenge is to move up one notch and to upgrade these cartoon models into real models. For example, it is, in our view, crucial to include in a consistent way the dynamics of trust -- and the possibility of sudden trust breakdown, along the lines sketched above -- in macro-economic models, in such a way that extreme events and crises can be 
accounted for. This is an ambitious project, which is only starting. \footnote{This is one of the ambitions of the CRISIS project, see: http://www.crisis-economics.eu/home}
 
Cartoon models are useful on many counts. But in the present case, did we learn anything that was not put by hand in the model? We believe so, at least insofar as  the models can be used to give  precise, well defined questions with possible empirical answers. For example, the robust signature of {\sc rfim}-like herding effects reported in section \ref{sect4A}, or the logarithmic spatial correlation of voting patterns (section \ref{sect4C}) are predictions of 
quantitative models  which were not expected a priori. Still, one would like to go beyond the level of plausibility and metaphors, and reach indisputable quantitative agreement between at least one non-trivial prediction of these self-referential feedback models and some empirical data. 
For example, the power-law distribution of avalanche sizes predicted by the {\sc rfim} is tantalizing and might enable one to understand
why bursts of activity and power-laws appear in e.g. the distribution of returns in financial markets. But there is not much more, at this stage, than wishful thinking and hand-waving. The possible existence of a condensation phase transition 
in multiple choice situations, of the type described in sections \ref{sect4D} and \ref{sect6A}, seems to be a highly relevant question, again amenable to precise empirical investigations (see e.g. \cite{Salganik,Kirman_fish}).

In our opinion, the most striking example of surprises (i.e. results that are not anticipated before a model is solved) encountered in this review is the clear {\it quantitative} demonstration, within a Schelling-type segregation model
\cite{Grauwin}, that Adam Smith's invisible hand can  fail badly at solving simple coordination problems -- see section \ref{sect5B}. We have also tried to insist on the issue of time-scales, that are 
generically extremely long in heterogenous interacting sytems (see e.g. \cite{Houches}). This may prevent a socially optimal equilibrium to be reached, unless some kind of catalysis nudges the system through an otherwise unsurmountable ``barrier'' -- see section \ref{sect2B}. As a theoretical question, the study of 
detailed-balance violating decision rules (which must be the rule rather than the exception) seems to me well worth the effort, in particular to decide whether conclusions based on detailed-balance rules are indeed robust and
generic. Of course, a unifying description of non-equilibrium stationary states is a major challenge in statistical physics as well! In any case, We hope the reader will be convinced that statistical physics can contribute to a better understanding of collective socio-economic phenomena, and that these theoretical ideas could one day be ripe enough to have a real impact on policy issues.  

\begin{acknowledgments} 

I thank J. Batista, S. Battiston, E. Beinhocker, E. Bertin, G. Biroli, Ch. Borghesi, M. Buchanan, D. Challet, R. Chicheportiche, R. Cont, P. Contucci, D. Farmer, P. Jensen, S. Grauwin, C. Hommes, A. Kirman, J. Kurchan, F. Lillo, M. Marsili, 
M. M\'ezard, Q. Michard, J.P. Nadal, A. Orl\'ean, L. Papaxanthos, M. Potters, J. Sethna, M. Tarzia, S. Wolf, M. Wyart, F. Zamponi and Y.C. Zhang for very useful conversations on these subjects over the
years, and S. Fortunato and S. Redner for giving me the opportunity to gather my thoughts on the subject. I thank in particular J. Batista, G. Biroli, D. Challet, A. Kirman, M. Marsili, D. Sornette, S. Wolf and the referees for carefully reading the manuscript and giving useful advice to improve it. This work is part of the European project 
CRISIS. I would like to dedicate this paper to Alan Kirman, whose work has been extremely inspiring to me. 

\end{acknowledgments}

\end{document}